\UseRawInputEncoding
\documentclass[10pt,aps,prb,final,twocolumn,amsmath,amssymb,groupedaddress,longbibliography]{revtex4-2}

\usepackage[pdftex,colorlinks=true,linkcolor=blue,citecolor=blue,urlcolor=blue]{hyperref}
\usepackage{dutchcal} % lower case caligraphic font
\usepackage{aurical}
\usepackage[T1]{fontenc}
\usepackage{units}
\usepackage{braket}
\usepackage{graphicx}
\usepackage{amsmath,amssymb}
\usepackage{amsfonts}
\usepackage{mathrsfs}
\usepackage{epsfig}
\usepackage{dcolumn}
\usepackage{enumerate}
\usepackage{amsthm}
\usepackage{bbold}
\usepackage{subfigure}
\usepackage{tikz}

\begin{document}

\title{A corrected Maslov index for complex saddle trajectories}

\author{Huichao Wang and Steven Tomsovic}
\affiliation{Department of Physics and Astronomy, Washington State University, Pullman, WA. USA 99164-2814}

\date{\today}

\begin{abstract}
Saddle point approximations, extremely important in a wide variety of physical contexts, require the analytical continuation of canonically conjugate quantities to complex variables in quantum mechanics.  An important component of this approximation's implementation is arriving at the phase correction attributable to caustics, which involves determinantal prefactors.  The common prescription of using the inverse of half a certain determinant's total accumulated phase sometimes leads to sign errors.  The root of this problem is traced to the zeros of the determinants at complex times crossing the real time axis.  Deformed complex time contours around the zeros can repair the sign errors that sometimes occur, but a much more practical way is given that links saddles back to associated real trajectories and avoids the necessity of locating the complex time zeros of the determinants.
\end{abstract}

\maketitle

\section{Introduction}
\label{intro}

Asymptotic analyses of linear wave equations give extremely powerful, quantitative methods of approximation in many fields of physics.  They are often called short wave or eikonal approximations for acoustic and electromagnetic waves~\cite{Pierce19, Born99}, where they enable an approximate construction of propagating waves or eigenmodes using rays.  In quantum mechanics, they often go by the names WKB, JWKB, EBK, semiclassical,  or saddle point methods ($\hbar\rightarrow 0$ or $N \rightarrow \infty$ for bosonic systems)~\cite{Einstein17, Jeffreys24, Wentzel26, Kramers26, Brillouin26, Keller58, Maslov81,Klauder85,Baranger01,Engl15}, where the approximation to eigenfunctions and the propagation of wave functions are built from classical trajectories.  

In the implementation of these methods for practical applications, perhaps the most subtle part is the phase adjustment due to encounters with caustics, where trajectories (or rays) pass through singularities in the prefactor that weights the contributions, i.e.~the inverse square root of a determinant that vanishes.  In the context of quantum mechanics, where the phase change is governed by what is typically called the Maslov index, a great deal has been written on this subject.  For example, there are notable works by Keller~\cite{Keller58}, Maslov and Fedoriuk~\cite{Maslov81}, Gutzwiller~\cite{Gutzwiller71}, and Littlejohn and co-authors~\cite{Creagh90}.  In the context of the stationary phase approximation applied to Feynman path integrals, which in its simplest form generates the Van Vleck-Gutzwiller propagator~\cite{Vanvleck28} approximation to the time-dependent Green function for the Schr\"odinger equation, Schulman connects the index to Morse theory~\cite{Schulman81}.  To paraphrase Keller~\cite{Keller58}:  `{\it the prefactor's phase is retarded by $m\pi/2$ on a trajectory which passes through a caustic on which the determinant vanishes to the $m^{th}$ order.'}  The Maslov index, $\nu$, is therefore just the sum of these m-integers for all of the caustics a trajectory passes through in its full history.  The end result is that the absolute value of the prefactor's inverse square root of a determinant gets multiplied by $\exp(-i\nu\pi/2)$, and the index can be considered to be defined modulus four~\cite{Littlejohn86}.

	Far less has been written about this issue for semiclassical approximations of quantum dynamics based on Glauber coherent states~\cite{Glauber63} or Gaussian wave packets~\cite{Littlejohn86, Huber88}, although they appear in an extremely wide variety of physical contexts~\cite{Scully97, Greiner02b, Polkovnikov11, Heller81b, Gruebele92, Zewail00, Agostini04}.  In the former, one is making a semiclassical approximation to a coherent state representation of a path integral~\cite{Klauder85, Baranger01}, and in the latter, one is carrying out the so-called generalized Gaussian wave packet dynamics (GGWPD) approximation, which is proven equivalent to a complex version of time-dependent WKB theory~\cite{Huber88}.  The underlying mathematics can be expressed in an essentially identical form for these two very different physical contexts.  Both necessitate a saddle point approximation instead of a stationary phase one and require complex trajectories.  For a typical saddle trajectory the determinant, being complex, never vanishes.  Instead, the phase just accumulates, and for the most part the correct phase for the prefactor is the inverse of half the determinant's total accumulated phase~\cite{Huber88, Baranger01}.  As such, there appears to be no need to discuss a Maslov index.  Nevertheless, it is helpful to translate the phase calculations into a Maslov index approach upon which later sections of this paper rely.  By calculating the determinant's total accumulated phase angle modulo $\pi$, its square root is brought into the first quadrant, call that phase $\theta$. The integer number of $\pi$ being dropped in the process becomes the index $\nu$, which gives a multiplying phase of $\exp(-i\nu\pi/2)$ for the prefactor multiplied by $\exp(-i\theta)$.  Implicit in this definition is that the modulus operation is defined to be the one that leaves a positive remainder even for negative numbers.  

It turns out that the half-the-total-accumulated-phase prescription sometimes gives an error in the phase index for complex trajectories, whether or not one is thinking in terms of a Maslov index, and therefore there is more to the full theory of the phase.  In the development of semiclassical coherent state dynamics in the context of the Bose-Hubbard model~\cite{Tomsovic18}, although not stated there, occasionally as a function of time, a saddle contribution would mysteriously and discontinuously change sign.  The purpose of this paper is to point out how to resolve this problem and give a corrected Maslov index for such cases.  

The structure of this paper is as follows.  The next section introduces notation, a bit of background information, such as the relevant determinants, and a very simple dynamical system used to illustrate the main ideas.  Section III discusses how the phase becomes incorrect and then shows how to repair the index, both in an ideal way and in another rather practical way.  This is followed by a short concluding section. 

\section{Background}
\label{bg}

It is helpful to give explicit expressions for the determinants that show up in the semiclassical approximation applied to Glauber coherent states and Gaussian wave packets.  The mathematics can be made to appear essentially identical through the application of quadratures, and repairing the Maslov index is the same in either context.  The discussion is presented for wave packets, but the main idea carries over immediately to coherent states.  The most interesting dynamical applications are to multi-degree-of-freedom systems, however, for the purposes of this paper, it is sufficient to illustrate the process with a simple one-degree-of-freedom example, which is taken to be the purely quartic oscillator.  The correction necessary extends immediately to multi-degree of freedom systems.

\subsection{Coherent states}
\label{cs}

A Glauber coherent state describing a bosonic many-body system takes the normalized form
\begin{equation}
\label{cse}
| {\bf z} \rangle = \exp \left(-\frac{\left| {\bf z}  \right|^2}{2} + {\bf z}  \hat a^\dagger \right)| {\bf 0}\rangle.
\end{equation}
Two quantities that are often useful are the overlaps of its evolved form $|{\bf z} (t)\rangle$ with some coherent state bra vector, $\langle {\bf z}^\prime | {\bf z} (t)\rangle$ (coherent state path integral matrix element), or projection onto quadrature variable bra vector eigenstates, $\langle {\bf x} | {\bf z} (t)\rangle$.  The saddle point approximation of either of these quantities results in a multiplication of the inverse square root of complex valued determinants.  Using quadratures, these quantities can be made to look identical to forms arising with Gaussian wave packets~\cite{Glauber63} whose determinants are given ahead; see~\cite{Wang21a}, the first of this pair of papers.  One discussion of the parameter mapping between them can be found in Appendix A of~\cite{Tomsovic18b}.  Roughly speaking, the complex parameters of ${\bf z}$ can be mapped onto pseudo-momentum and -position centroids, and the ground state links the shape parameters.  

\subsection{Gaussian wave packets}
\label{wp}

A normalized, multi-dimensional Gaussian wave packet may be parameterized as follows ($\alpha$ labels the entire set):
\begin{eqnarray}
\label{wavepacket}
&&\phi_\alpha(\vec x) = \nonumber \\
&&{\cal N}_\alpha^0\exp\left[ - \left(\vec x - \vec q_\alpha \right) \cdot \frac{{\bf b}_\alpha}{2\hbar} \cdot \left(\vec x - \vec q_\alpha \right) +\frac{i \vec p_\alpha}{\hbar} \cdot \left(\vec x - \vec q_\alpha \right)\right], \nonumber \\
\end{eqnarray}
where
\begin{equation}
\label{wavepacket2}
 {\cal N}_\alpha^0 = \left[\frac{{\rm Det}\left({\bf b}_\alpha+{\bf b}^*_\alpha\right)}{(2\pi\hbar)^N}\right]^{1/4}\exp \left( \frac{i}{2\hbar}\vec p_\alpha \cdot \vec q_\alpha \right). 
\end{equation}
The subscript $\alpha$ is a label for the parameters that define the particular wave packet, $\vec x$ is the position variable for the quantum system, and $(\vec q,\vec p)$ are the canonically conjugate position-momentum phase space variables for the analogous classical system in ordinary Schr\"odinger quantum mechanics of single or few particles.  The phase convention chosen here corresponds exactly to the coherent state of Eq.~\eqref{cse} projected onto a quadrature variable.  The real centroid is given by $(\vec q_\alpha, \vec p_\alpha)$ and the variances and covariances determined by the symmetric matrix ${\bf b}_\alpha$ (if ${\bf b}_\alpha$ is complex, the wave packet has a chirped phase dependence).  This form has the advantage that $\hbar$ does not explicitly appear in the equations that define the saddle's two point boundary conditions. 

It also leaves the overall shape of its Wigner transform independent of $\hbar$, other than the volume (overall scale).  The result is
\begin{equation*}
{\cal W}(\vec p, \vec q) = \frac{1}{(2\pi\hbar)^{N}} \int_{-\infty}^\infty {\rm d} \vec x \ {\rm e}^{i \vec p \cdot \vec x/\hbar} \phi_\alpha \left(q-\frac{\vec x}{2}\right)  \phi^*_\alpha \left(q+\frac{\vec x}{2}\right) 
\end{equation*}
\begin{equation}
\label{wignerwp}
= \left(\pi \hbar \right)^{-N} \exp \left[ - \left(\vec p - \vec p_\alpha, \vec q - \vec q_\alpha \right) \cdot \frac{{\bf A}_\alpha}{\hbar} \cdot \left(\vec p - \vec p_\alpha, \vec q - \vec q_\alpha \right) \right], 
\end{equation}
where ${\bf A}_\alpha$ is
\begin{equation}
\label{mvg}
{\bf A}_\alpha = \left(\begin{array}{cc}
{\bf  c^{-1}} & {\bf  c}^{-1} \cdot {\bf  d}  \\
 {\bf  d} \cdot {\bf c}^{-1} & {\bf c} + {\bf  d} \cdot {\bf c}^{-1} \cdot {\bf  d} \end{array}  
\right),   \qquad {\rm Det}\left[ {\bf A}_\alpha \right] =1
\end{equation}
with the association 
\begin{equation}
\label{mvgwf}
{\bf b}_\alpha = {\bf c} + i {\bf d}.
\end{equation}
The $ 2N \times 2N$ dimensional matrix ${\bf A}_\alpha$  determines the shape parameters; $\hbar$ is seen to enter only as a scale in Eq.~\eqref{wignerwp}.  ${\bf A}_\alpha$ is real and symmetric.  If $\bf b_\alpha$ is real, there are no covariances between $\vec p$ and $\vec q$.  The off-diagonal blocks of the matrix ${\bf A}_\alpha$ disappear.  

Note that a matrix element of the coherent state path integral could be expressed in quadratures as
\begin{equation}
\label{ampc}
{\cal A}_{\beta\alpha}(t) = \int_{-\infty}^\infty {\rm d}\vec x\ \phi^*_\beta(\vec x) \phi_\alpha(\vec x;t)
\end{equation}
and which could be thought of as a transport coefficient for wave packets.  If $\beta=\alpha$, it would be a diagonal element or a return amplitude.

\subsection{The determinants of interest}
\label{gwp1}

The saddle trajectories are found using Hamiltonians analytically continued for complex canonically conjugate variables denoted with a calligraphic font $({\cal p},{\cal q})$.  Using the definition of a trajectory's stability matrix
\begin{equation}
\left( \begin{array}{c} \delta \vec {\cal p}_t \\ \delta \vec {\cal q}_t \end{array} \right) =  \left( \begin{array}{c} \bf{M_{11}} \\ \bf{M_{21}} \end{array} \begin{array}{c} \bf{M_{12}} \\ \bf{M_{22}} \end{array} \right)
\left( \begin{array}{c} \delta \vec {\cal p}_0 \\ \delta \vec {\cal q}_0 \end{array} \right), 
\label{delta}
\end{equation}
the determinants of interest are as follows.  For the semiclassical approximation to the time-dependent Green function (the van Vleck-Gutzwiller propagator), the van Vleck determinant is given by
\begin{equation}
\label{det0}
{\cal D}_0 =  {\rm Det}\left({\bf -M_{21}}\right),
\end{equation}
where the propagator has a multiplicative factor ${\cal D}_0^{-1/2}$ and the trajectory used to generate the stability matrix is real as well as ${\cal D}_0$.  For propagating a wave packet or coherent state, the appropriate determinant becomes
\begin{equation}
\label{det1}
{\cal D}_1 =  {\rm Det}\left({\bf M_{22}} + i {\bf M_{21}}\cdot {\bf b}_\alpha  \right)
\end{equation}
 and the trajectory used is almost always complex, as is the determinant ${\cal D}_1$.  Similarly, for the semiclassical approximation to a coherent state path integral matrix element or wave packet defined transport coefficient, the determinant turns out to be
\begin{align}
\label{det2}
&{\cal D}_2  =  \nonumber \\
&{\rm Det}\left[{\bf M_{11}}\cdot {\bf b}_\alpha + {\bf b}^*_\beta \cdot {\bf M_{22}} + i \left({\bf b}^*_\beta\cdot  {\bf M_{21}}\cdot {\bf b}_\alpha - {\bf M_{12}} \right) \right]
\end{align}
which is again complex in general.  The phase index of concern in this paper originates from either ${\cal D}_1^{-1/2}$ or ${\cal D}_2^{-1/2}$.

\subsection{The purely quartic oscillator}
\label{quarticoscillator}

The one degree of freedom purely quartic oscillator has a number of simple features that makes it ideal for illustrating how to correct the phase index for complex determinants.  Setting $\hbar=m=1$, its Schr\"odinger equation is given by
\begin{equation}
 i\frac{\partial}{\partial t} \phi(x;t) = \left(-\frac{\partial^2}{2\partial x^2} + \lambda x^4 \right) \phi(x;t).
\label{schroedinger}
\end{equation}
The corresponding analytically continued Hamiltonian takes the form [complex (${\cal p}, {\cal q}$)] \begin{equation}
\label{quosc}
H( {\cal p}, {\cal q}) = \frac{{\cal p}^2}{2} + \lambda {\cal q}^4,
\end{equation}
where $\lambda= 0.05$ is the value taken for all the illustrations shown in this paper.  Being homogeneous, there are simplifying scaling relations between trajectories on different energy surfaces~\cite{Bohigas93}.  They provide convenient checks on various calculations.

The potential generates nonlinear Hamiltonian equations that lead to sufficiently complicated behaviors for our purposes.  There is a shearing behavior in the dynamics transverse to the energy surface, and at any given time, there is an infinity of saddle solutions~\cite{Wang21a}, only a handful of which are physically relevant (neither on the wrong side of Stokes lines nor with imaginary classical actions so large as to contribute negligibly).  Here, the focus is exclusively on the physically relevant saddle trajectories.

\section{Repairing the Maslov index}
\label{repair}

The need to look deeper into the phase index was first realized by a problem that arose in developing the semiclassical method for Glauber coherent states governed by the Bose-Hubbard Hamiltonian~\cite{Tomsovic18}.  The problem was then reproduced in a different way with the much simpler quartic oscillator, where it was easier to analyze fully.  We start by illustrating the issue.

\subsection{What goes wrong?}
\label{wrong}

\begin{figure}[htb]
\includegraphics[width=8.5 cm]{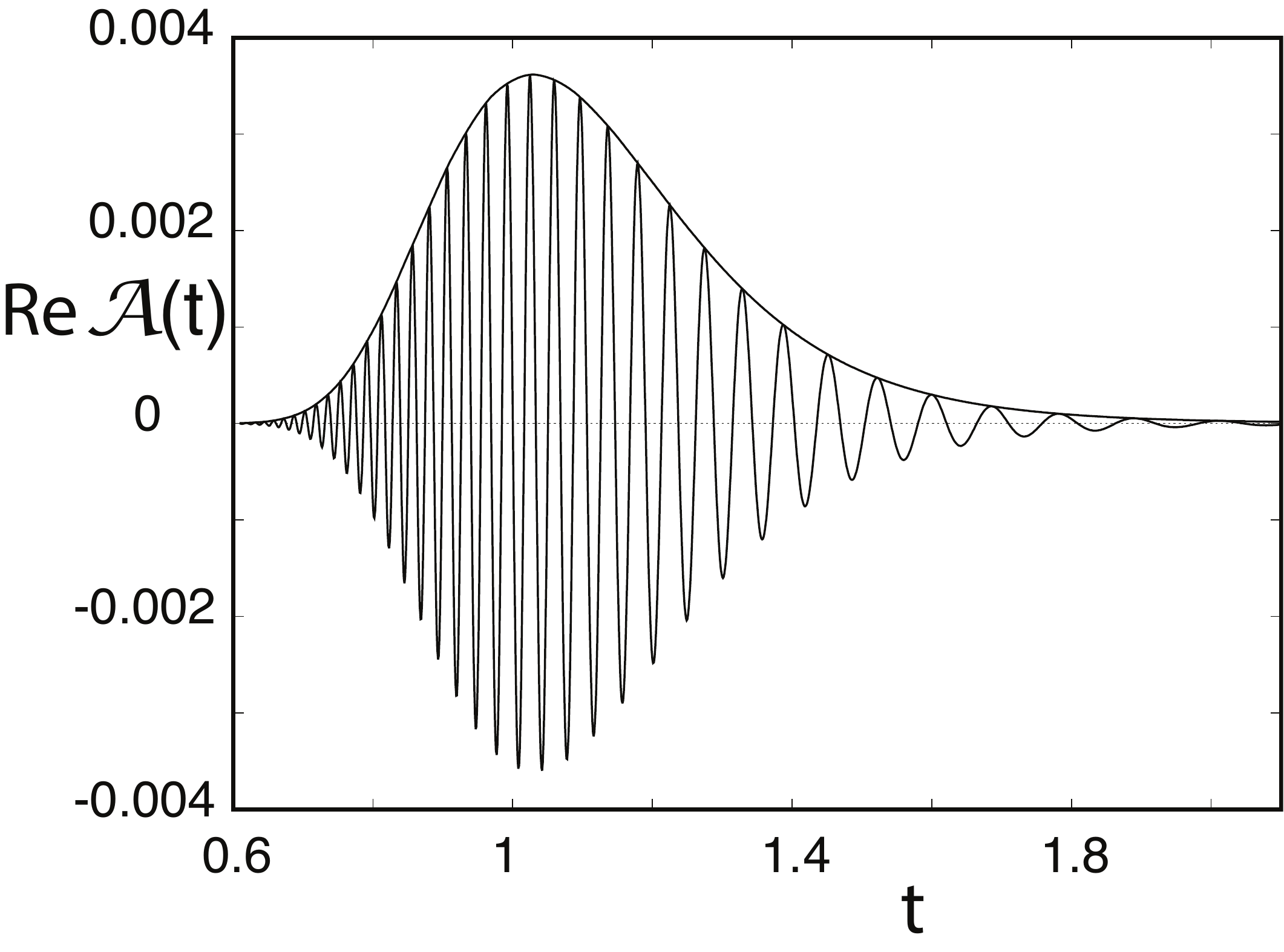} 
\caption{The real part of ${\cal A}(t)$ for one particular saddle family; the envelope is the absolute value.  There is a faster phase oscillation at short times decreasing as time increases corresponding to changes in the complex saddle trajectory.   At short times, the real parts of the saddle trajectory's energy are greater than the wave packet's energy expectation value, and at longer times they are less.  The saddle family's peak contribution occurs near where the two are equal.  
 \label{fig1}} 
\end{figure}

The issue presented below is a general problem that can arise whatever the number of degrees of freedom or the nature of the dynamics, be it regular or chaotic.  Consider a transport coefficient, ${\cal A}_{\beta\alpha}(t)$, with the parameter set $\beta=\alpha$ and drop the subscript in the notation; this involves a ${\cal D}_2$.  For continuous time dynamical systems, each saddle found at some given time gives rise to a one parameter family of saddles labeled by time.  As $t$ changes continuously, the saddle trajectory's initial conditions change continuously as well.  Barring orbit bifurcations and crossing Stokes surfaces (which becomes exceedingly unlikely in the $\hbar\rightarrow 0$ limit), this one parameter family generates a continuous contribution to ${\cal A}(t)$.  A typical example was given in~\cite{Tomsovic18b}, and is reproduced here; see Fig.~\ref{fig1}.  
Generally speaking, for an ${\cal A}(t)$ there is a peak contribution time for a saddle family corresponding to a saddle trajectory possessing an energy close to that of the wave packet.  Earlier and later in time, the saddle trajectory moves further away from this energy and the contribution decays, thus creating a time window in which it contributes significantly.  

In~\cite{Tomsovic18} for the Bose-Hubbard model, the majority of the saddle families produced contributions to ${\cal A}(t)$ that were continuous and well-behaved as in the example of Fig.~\ref{fig1}.  However, sometimes a family's contribution was observed to change sign discontinuously at some particular time, $t_0$, i.e.~acquires an unnecessary extra phase $\exp(\pm i\pi)$ (or $\nu$ an extraneous $\pm 2$).  Discontinuities are unphysical and require correction.  We emphasize that for each particular time, $t$, there is a unique saddle trajectory that has its own evolution from $0$ to $t$.  The saddle trajectory for $t_1=t_0-\delta t$ has its determinant's phase followed continuously up to $t_1$.  Likewise, the saddle trajectory for $t_2=t_0+\delta t$ has its determinant's phase also followed continuously, and yet there is a discontinuous jump between the two.  

\begin{figure}[htb]
\includegraphics[width=8.5 cm]{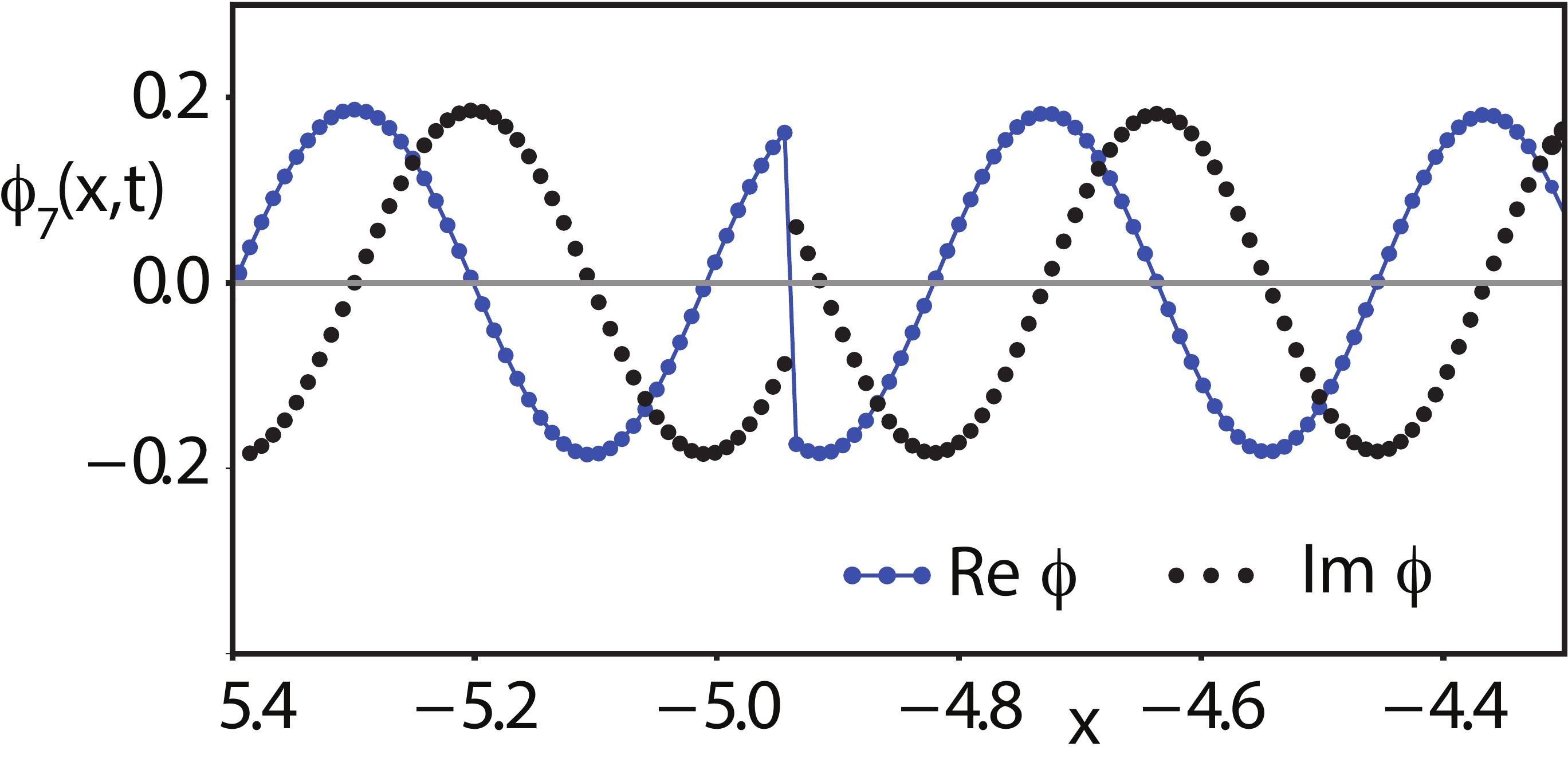} 
\caption{Contribution to the evolution of a wave packet from a single continuous family of saddles as a function of position; the saddles are linked to a classical foliation of real initial conditions labelled by \textcircled{\raisebox{-0.9pt}{7}} in Fig.~\ref{fig5} ahead.  The initial wave packet, Eq.~\eqref{wavepacket}, is centered at $(q_\alpha,p_\alpha)= (0,20)$ of width parameter $b_\alpha=32$ and propagated with Eq.~\eqref{schroedinger}.  The propagation time is equal to $3\tau$ where $\tau$ is the period of motion for the centroid initial condition $(q_\alpha,p_\alpha)$.
 \label{fig2}} 
\end{figure}

To track down what causes the discontinuous sign flip, it is simpler to consider an evolved wave packet $\phi(x,t)$ for the quartic oscillator.  The prefactor involves a ${\cal D}_1$, and an example of a sign discontinuity is shown in Fig~\ref{fig2}.
The Maslov index is created by continuously following the total phase accumulation for each saddle individually modulo $\pi$ and counting the integer number of $\pi$ dropped in the process as described in the introduction.  Despite following the phase accumulation continuously, the index is off by $2$ (modulo $4$) for all the saddles to the left of the abrupt sign flip in one saddle family's contributions to $\phi(x,t)$ labelled with the subscript \textcircled{\raisebox{-0.9pt}{7}}, i.e. $\phi_{\footnotesize{\textcircled{\raisebox{-0.9pt}{7}}}}(x,t)$.

\begin{figure}[htb]
\centering
\begin{subfigure}
\centering
\includegraphics[width=4 cm]{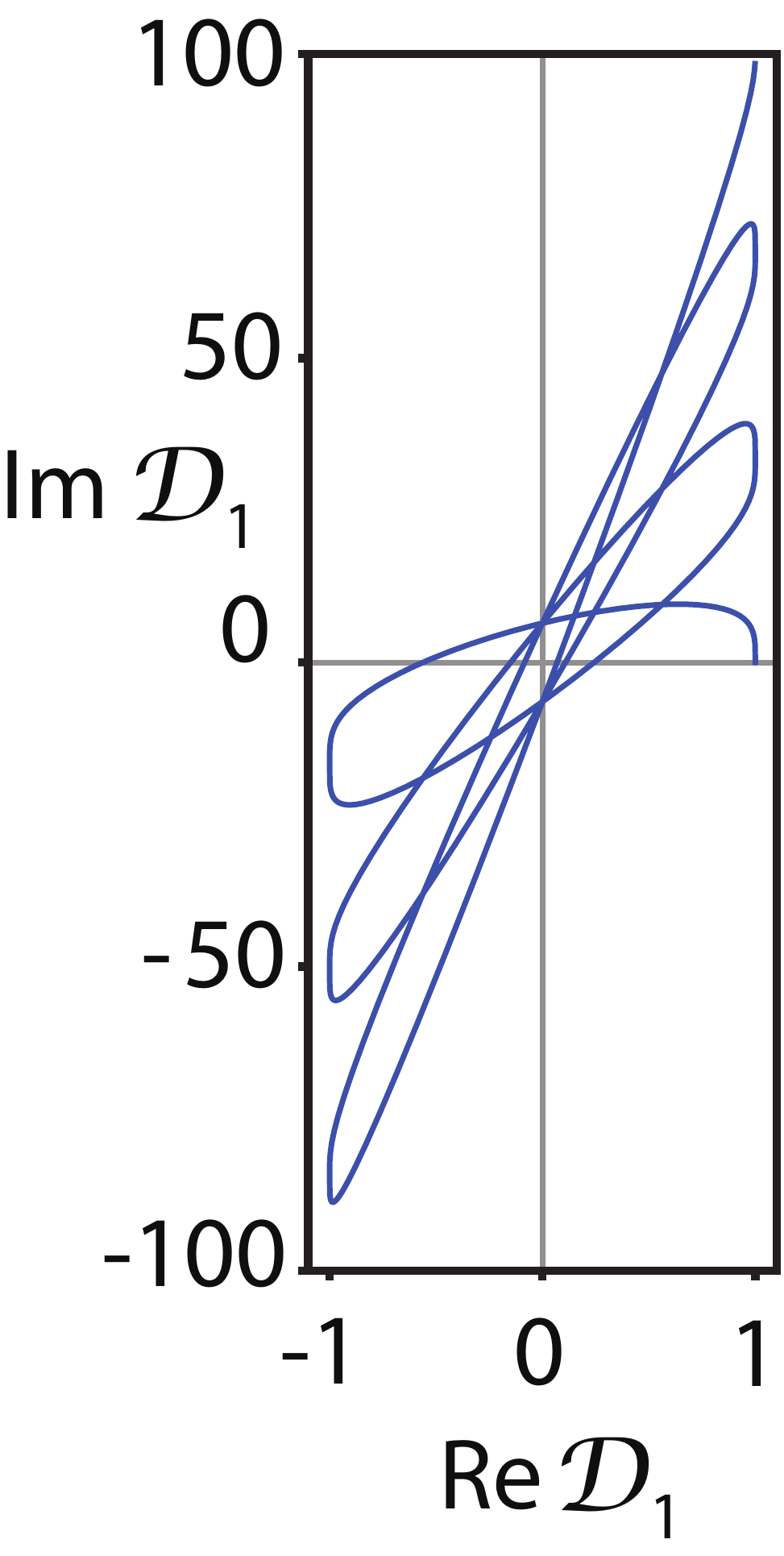} 
\label{fig3a}
\end{subfigure}
\begin{subfigure}
\centering
\includegraphics[width=4 cm]{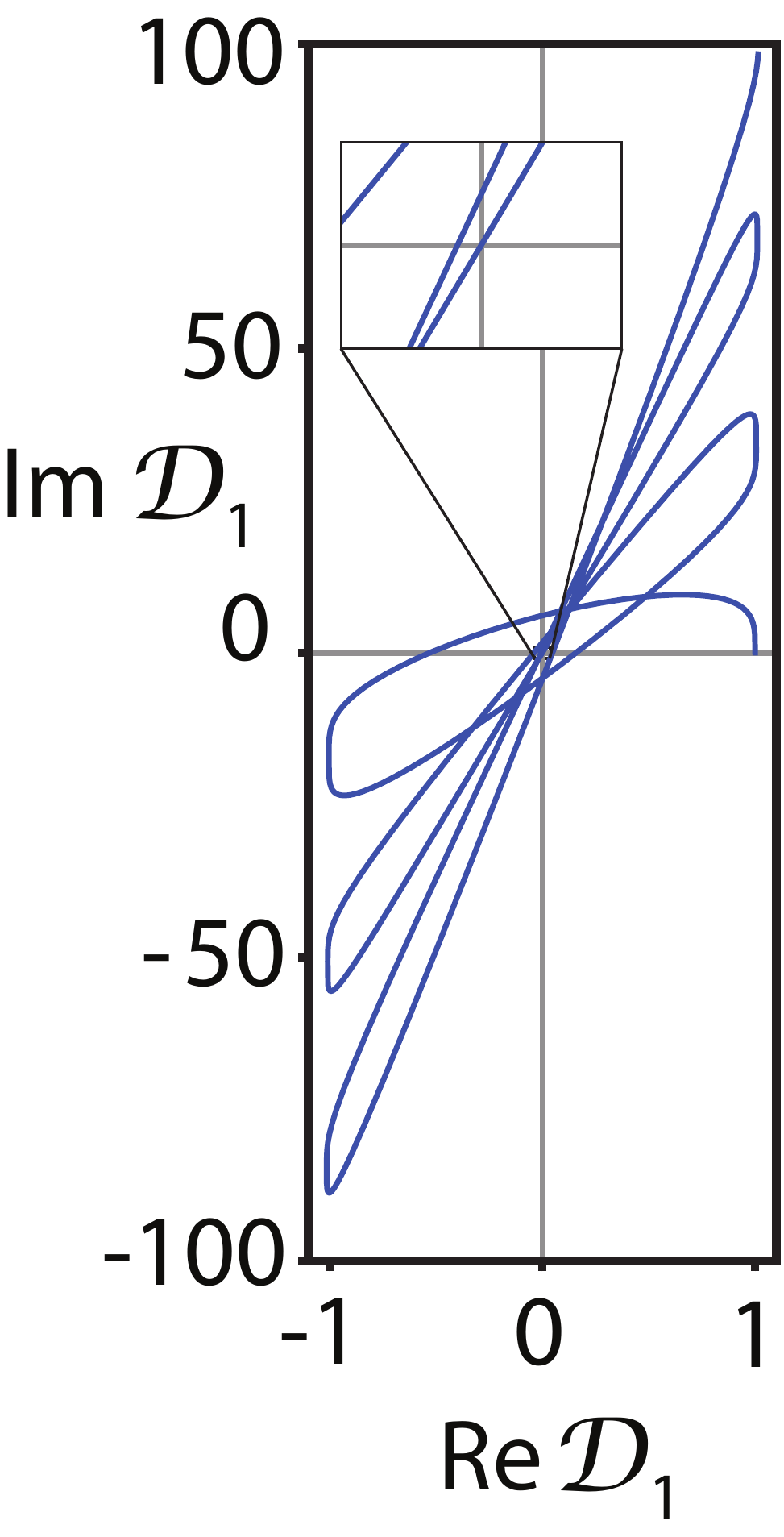}
\label{fig3b}
\end{subfigure}
\caption{The evolution of the total accumulated phase for a time $t=3\tau$.  In the left panel, the real and imaginary parts of ${\cal D}_1$  are plotted for all times $0\le t \le 3\tau$.  Real trajectory initial conditions are used.  At $t=0$, the ${\cal D}_1$ curve starts at the point (1,0) and proceeds always in a counterclockwise direction as time increases up to the final time $3\tau$.  In the right panel, the phase for the critical saddle begins similarly, but the complexification of the stability matrix elements generates a point at which ${\cal D}_1=0$.  
 \label{fig3}} 
\end{figure}

The determinant of interest, ${\cal D}_1$, for any real trajectory has a phase that evolves in time by rotating counterclockwise around zero.  See the illustration of this in the left panel of Fig.~\ref{fig3}, which follows the 
phase accumulation for a real initial condition up to a time $t=3\tau$.  However, there is a {\bf critical} complex saddle trajectory associated with the discontinuity in Fig.~\ref{fig2} for which ${\cal D}_1 = 0$ at some time during its past; see the right panel of Fig.~\ref{fig3}.  This creates an uncertainty about its phase accumulation.  For saddles to the left of this critical saddle in Fig.~\ref{fig2} the total phase accumulation curve passes to the left of zero in the neighborhood of where ${\cal D}_1$ almost vanishes, whereas for those saddles to the right of the critical saddle, the phase curve passes on the right side of zero.  The saddles just to the left acquire $2\pi$ less total phase than those just to the right.  Hence, the discontinuous sign flip in the semiclassical contribution labelled $\phi_{\footnotesize{\textcircled{\raisebox{-0.9pt}{7}}}}(x,t)$.

\subsection{Complex time paths}
\label{repairctp}

Generally speaking, the zeros of ${\cal D}_1$ are found at complex values of the time $t$.  This is illustrated in \begin{figure}[htb]
\includegraphics[width=8.5 cm]{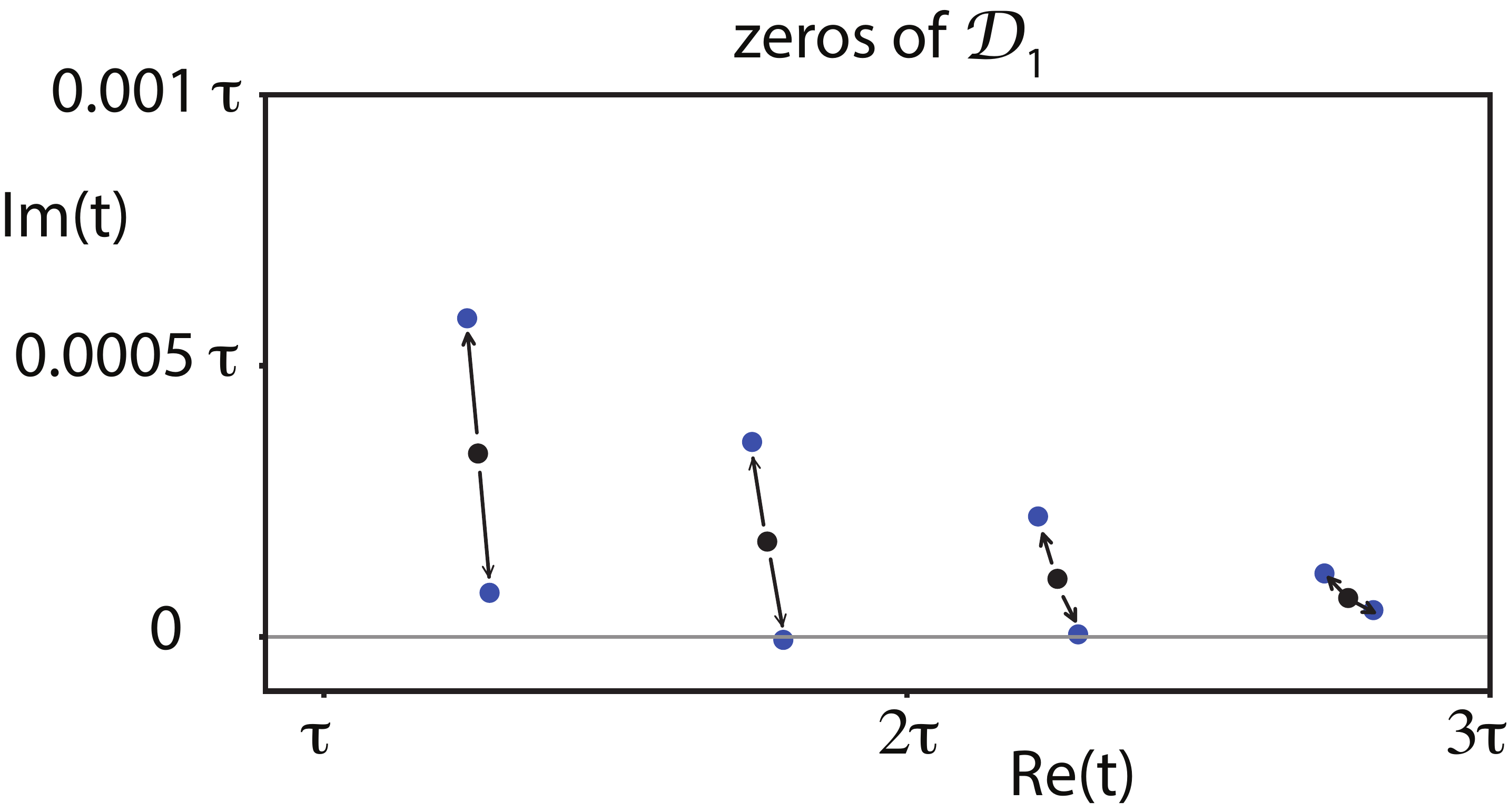} 
\caption{The location in time of zeros for $3$ saddles from foliation $\textcircled{\raisebox{-0.9pt}{7}}$.  The middle zero at each of the arrows' origination point is for $x=0.0$.  Following the up arrows are the zeros for $x=1.9008$, and likewise, following the down arrows $x=-1.9008$.  For this latter point, the zero a bit before $t=2\tau$ has just crossed the real axis, the point $x=-1.9008$ is just to the left of a sign flip.  The next zero is very close to crossing, but has not yet.  A small further decrease in the $x$-value leads to a second sign flip.  There is one zero not pictured for $0\le t\le \tau$, but its imaginary value is quite large and positive too high to be seen on the figure's scale.
 \label{fig4}} 
\end{figure}
Fig.~\ref{fig4}.  The location of the zeros are shown for $3$ values of the position in the wave function, $x\approx -1.9,\ 0,\ 1.9$.  For the point $x=0.0$, this example was designed so that the saddle trajectory's initial conditions coincide with the wave packet centroid $(q_\alpha,p_\alpha)$ since $t=3\tau$ and $q_\alpha=0$.  Therefore, its ${\cal D}_1$ wraps exclusively counterclockwise around zero exactly as in the left panel of Fig.~\ref{fig3}.  The zeros of its ${\cal D}_1$ are all found above the $Im(t)=0$ axis and are the central points between arrows.    The total accumulated phase always increases and, hence, the Maslov index never drops with increasing time.  To locate the zeros for $x\approx 1.9$ follow the up arrows; i.e.~the zeros have all moved upward and the Maslov index for the $x\approx 1.9$ saddle also always increases exactly like the $x=0.0$ case.  

However, the zeros for the $x\approx -1.9$ case all drop and are found at the end of the down arrows.  Just before $t=2\tau$, one of its zeros has crossed the $Im(t)=0$ axis.  To the left and right of $x\approx -1.9$, there is a discontinuous change of $2\pi$ in the total phase accumulation (for a saddle of $x > -1.9$, there is $2\pi$ greater phase accumulation than for $x < -1.9$).   Thus, there is a discontinuous sign change in the saddle family contributions to the propagated wave function exactly as pictured in Fig.~\ref{fig2}, and the Maslov index drops by 2 as $x$ decreases across the discontinuity.  Therefore, it is possible to end up on the wrong branch of the square root of the relevant determinant following a real time propagation path if one is just accumulating the total phase and taking half the inverse.  This is rectified by deforming the $t$ integration contour in this neighborhood to negative imaginary times such that the contour runs beneath the zero that has crossed the $Im(t)=0$ axis.  Knowledge of the location of all the pertinent zeros and appropriately deforming the time contour leads to the corrected Maslov index.

\subsection{A practical approach}
\label{practical}

It is quite impractical, especially for multi-degree-of-freedom systems, to be required to develop complex time evolution and locate the zeros of the determinant of interest for all the saddles.  For physically relevant saddles, there are ways of avoiding complex time entirely.  We give one such rather practical prescription here that relies on continuity and parameter variation.  This prescription depends on a classification of the saddles into two groups, exposed and hidden saddles~\cite{Wang21a}.  The exposed saddles correspond to classically allowed processes, and the hidden saddles correspond to classically non-allowed processes, such as tunneling.

\subsubsection{Exposed saddles}

Basically, exposed saddles can be located by using Newton-Raphson search schemes whose initial seeds are real trajectories' initial conditions corresponding to distinct classically allowed pathways (transport)~\cite{Vanvoorhis02,Vanvoorhis03,Pal16,Tomsovic18b,Oconnor92,Tomsovic93,Tomsovic93b, Barnes94}.  The Wigner transform is especially useful for this purpose~\cite{Wang21a}.  For the initial wave packet example defined in the Fig.~\ref{fig2} caption, the Wigner transform is illustrated in 
\begin{figure}[htb]
\includegraphics[width=8.5 cm]{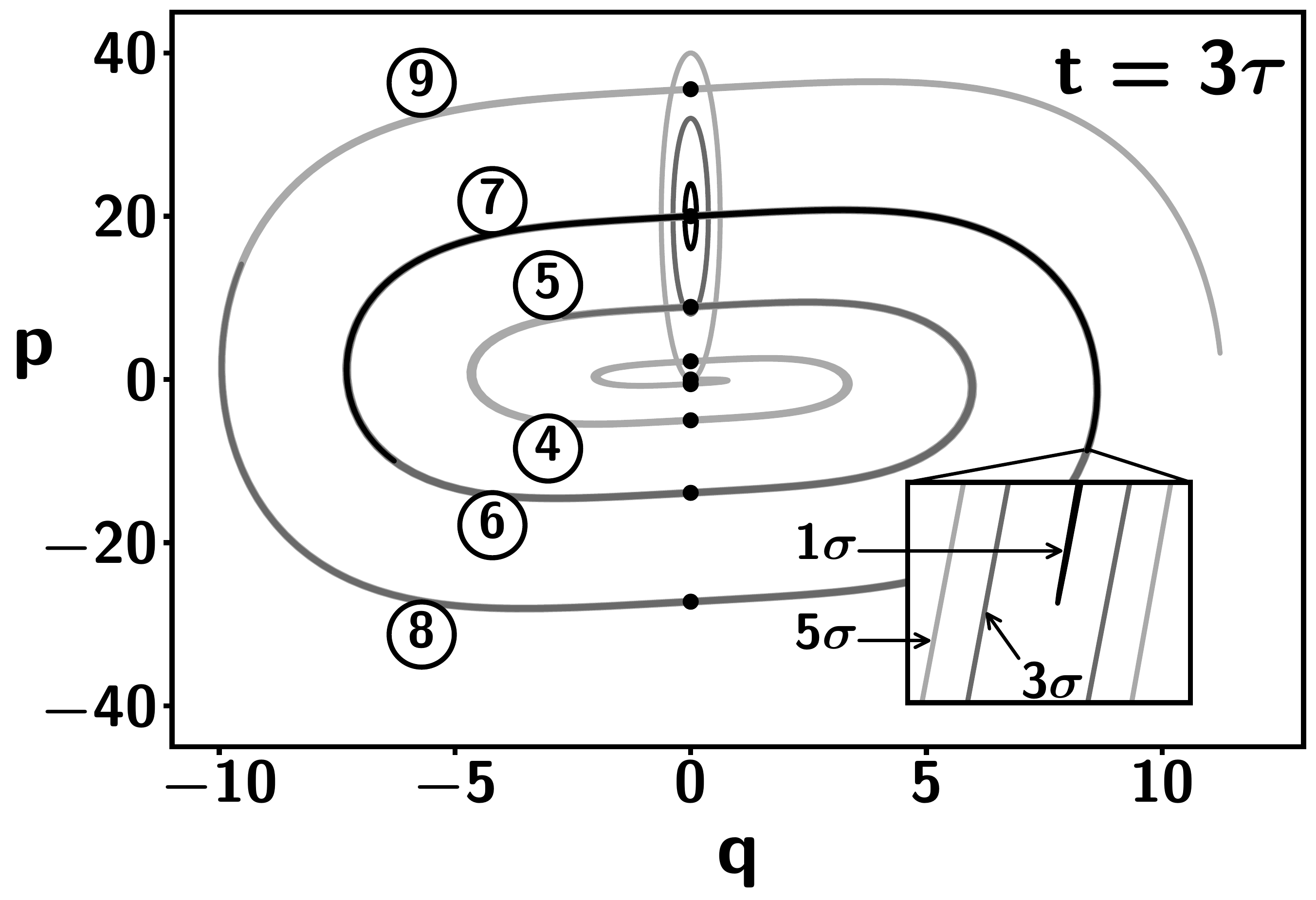}
\caption{Illustration of distinct classical transport pathways.  The vertically oriented ellipses are contours of the Wigner transform of the initial wave packet defined using a single degree of freedom version of Eq.~\eqref{wavepacket} and the parameters given in the Fig.~\ref{fig2} caption.  They are propagated classically for $3$ periods of motion of the centroid, and they generate the distorted spirals.  The $5\sigma$ contour can be divided into $9$ foliations, a trajectory member of which can be used to generate initial conditions for a Newton-Raphson scheme to locate unique saddles.  The blackened dots show the locations of the real parts of the $9$ associated saddles' complex final phase space points contributing to $\phi(x=0,t)$ (the initial conditions follow by back propagating a time $3\tau$).  
 \label{fig5}} 
\end{figure}
Fig.~\ref{fig5} by showing the ellipses corresponding to the $\sigma,\ 3\sigma,\ 5\sigma$ contours of the resulting Gaussian functional form.  These ellipses are propagated for a time $3\tau$ and result in the collapsed distorted spirals also plotted.  Consider the $5\sigma$ spiral (the most extended one).  It can be subdivided into $9$ foliations or groups of classical trajectories that are somewhat similar; for example, foliation \textcircled{\raisebox{-0.9pt}{7}} contains the central orbit whose initial condition is $(q_\alpha,p_\alpha)$. It consists of other trajectories that have similarly made approximately $3$ full cycles of the motion, whereas foliation \textcircled{\raisebox{-0.9pt}{9}} contains orbits that have completed approximately $4$ full cycles of the motion.  The boundaries between the foliations are best located at the classical turning points of the motion for this case.  Each foliation can be used to generate a unique saddle at a given point $x$ by a Newton-Raphson search scheme, and therefore a continuous family of saddles (as a function of $x$) can be labelled the same as the foliation.

The key idea is that there must be consistency in sign between the result for a saddle associated with some foliation and the sign one would arrive at by making use of the real trajectories of the same foliation.  In other words, the Van Vleck-Gutzwiller propagator relying on real trajectories of that foliation could be used to evaluate a contribution to, say an ${\cal A}_{\beta\alpha}(t)$ or $\phi(x,t)$, and that must give a consistent result with using the true saddle approximation relying on complex saddle trajectories.  

There are many reasons to expect this consistency, one of which is that shrinking the width of the wave packet makes it approach a position eigenstate in a continuous way and $\phi(x,t)\rightarrow G(x,q_\alpha;t)$ in the limit of vanishing width.  Put more simply, the stability matrix of the associated real trajectory relied upon to locate the exposed saddle in the Newton-Raphson search scheme could be used to calculate the total phase accumulation of ${\cal D}_1$ or ${\cal D}_2$, and thereby deduce the corrected Maslov index.  

\begin{figure}
    \begin{subfigure}
            \centering
            \includegraphics[width=8.3 cm]{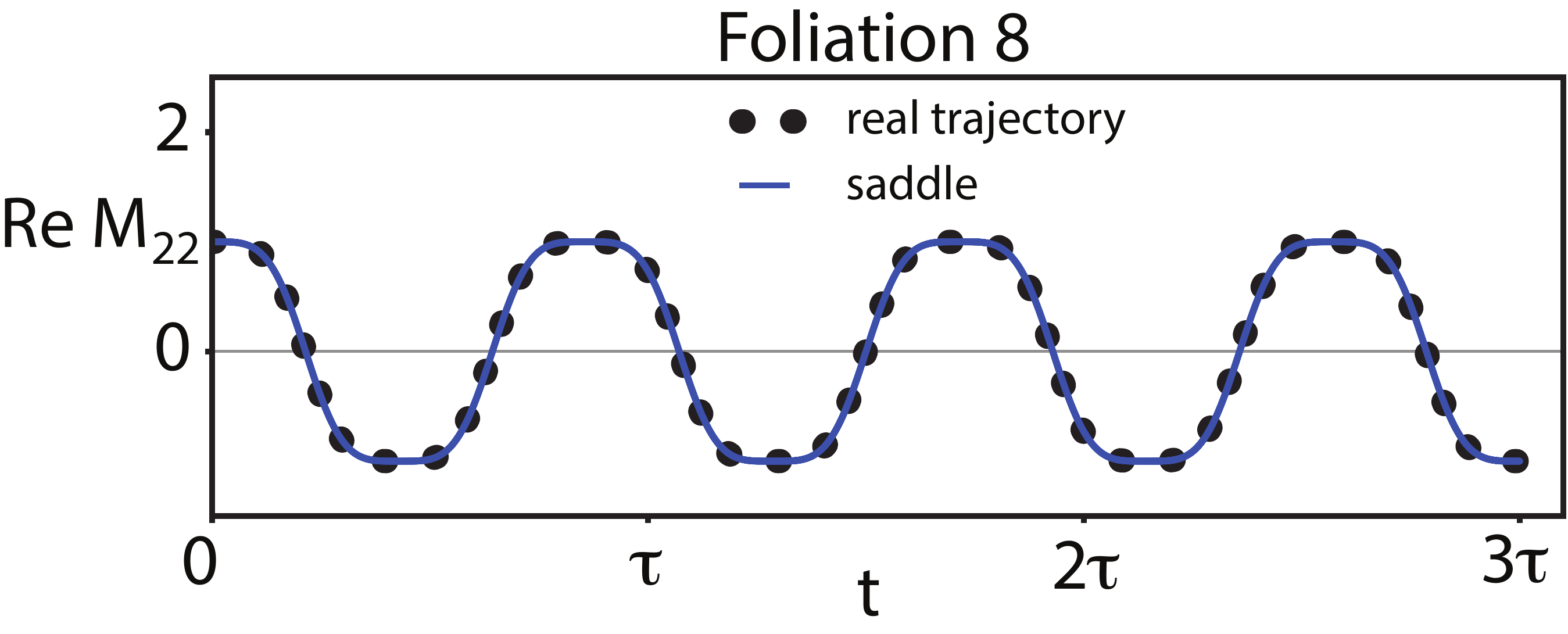}
            \label{fig6a}
    \end{subfigure}
    \begin{subfigure}%[b]{0.5\textwidth}
            \centering
            \includegraphics[width=8.3 cm]{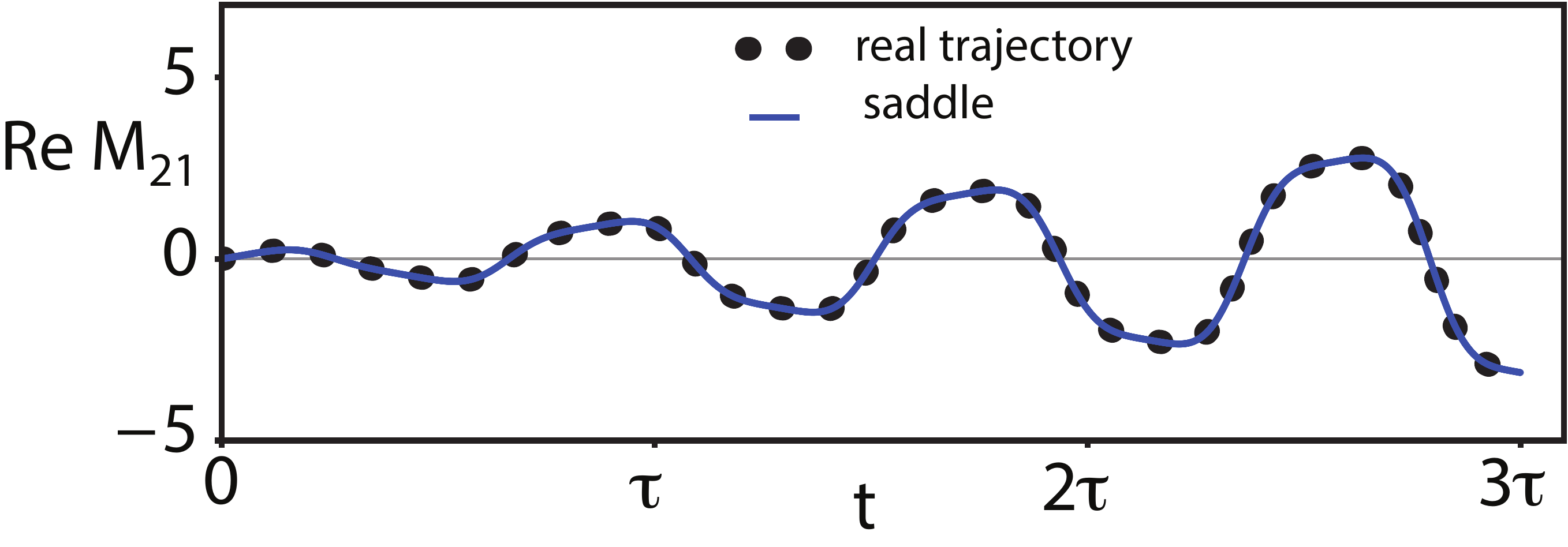}
            \label{fig6b}
    \end{subfigure}
    \caption{The real parts of $M_{22}$ and $M_{21}$ for the saddle of foliation \textcircled{\raisebox{-0.9pt}{8}} at $x=0$ and its associated real trajectory.  If the stability matrix elements were not so similar, the Newton-Raphson search for the saddle could not converge since they are essential to the method.}
\label{fig6}
\end{figure}

\begin{figure}
    \begin{subfigure}
            \centering
            \includegraphics[width=8.3 cm]{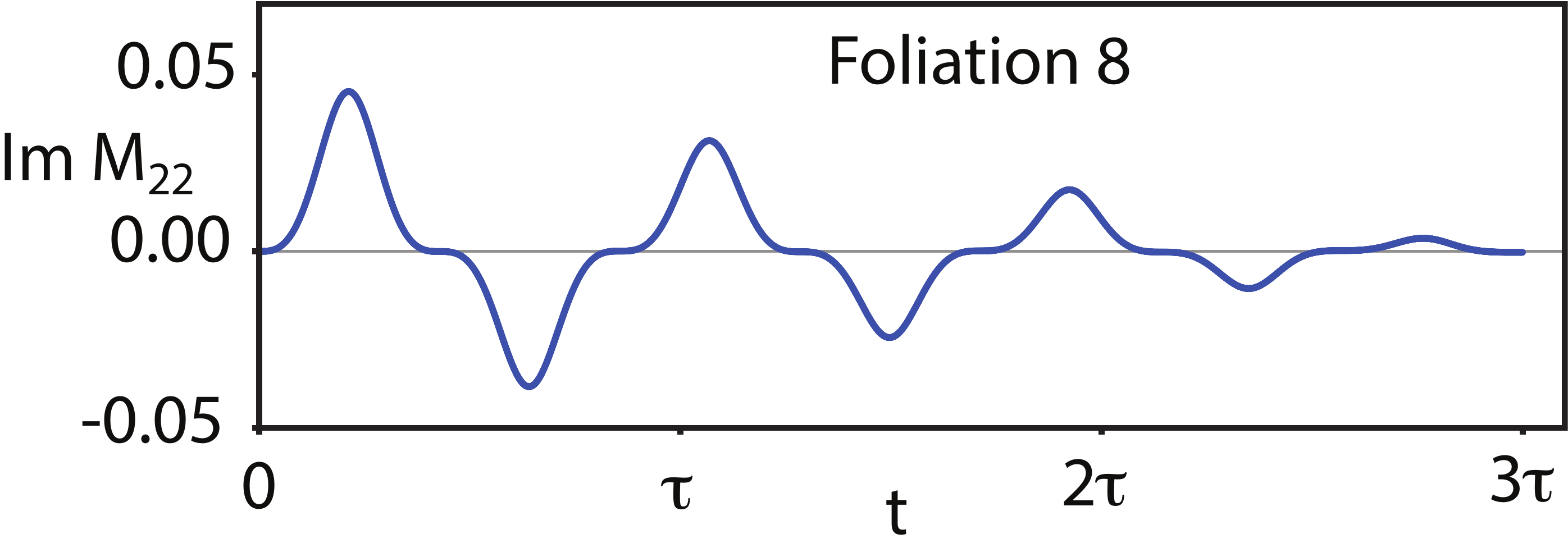}
            \label{fig7a}
    \end{subfigure}
    \begin{subfigure}%[b]{0.5\textwidth}
            \centering
            \includegraphics[width=8.3 cm]{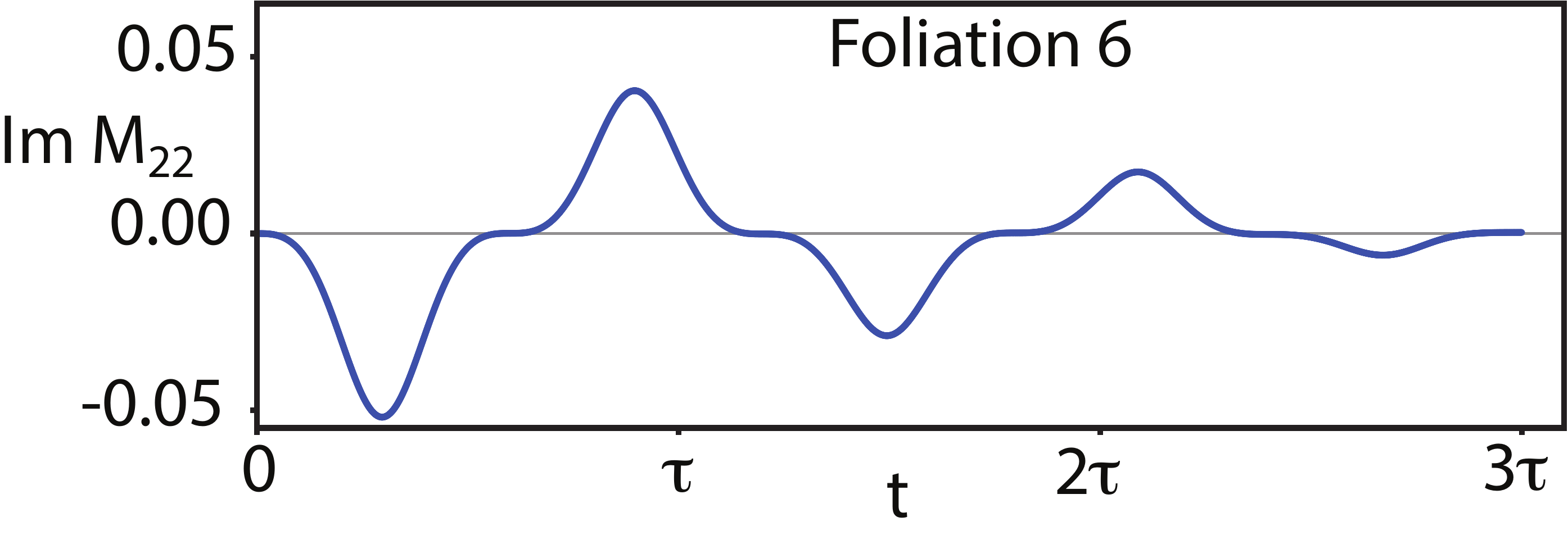}
            \label{fig7b}
    \end{subfigure}
    \caption{The imaginary parts of $M_{22}$ for the saddles of foliations \textcircled{\raisebox{-0.9pt}{8}} and \textcircled{\raisebox{-0.9pt}{6}} at $x=0$.  For the associated real trajectories, the imaginary parts vanish for all times.  The stability matrix elements' imaginary parts using the saddles are orders of magnitude smaller than the real parts of Fig.~\ref{fig6}.  As noted the saddle of foliation \textcircled{\raisebox{-0.9pt}{7}} is equivalent to its associated real orbit, and its imaginary parts have vanished for this case.  Foliations \textcircled{\raisebox{-0.9pt}{8}} and \textcircled{\raisebox{-0.9pt}{6}} are the closest neighboring foliations on opposite sides.  Their stability matrix elements' imaginary parts are reflected, which is a trend that continues.  All the lower energy foliations possess an $Im(M_{11})$ that goes negative first, which is the opposite of the higher energy foliations.}
\label{fig7}
\end{figure}

\begin{figure}[htb]
\includegraphics[width=8 cm]{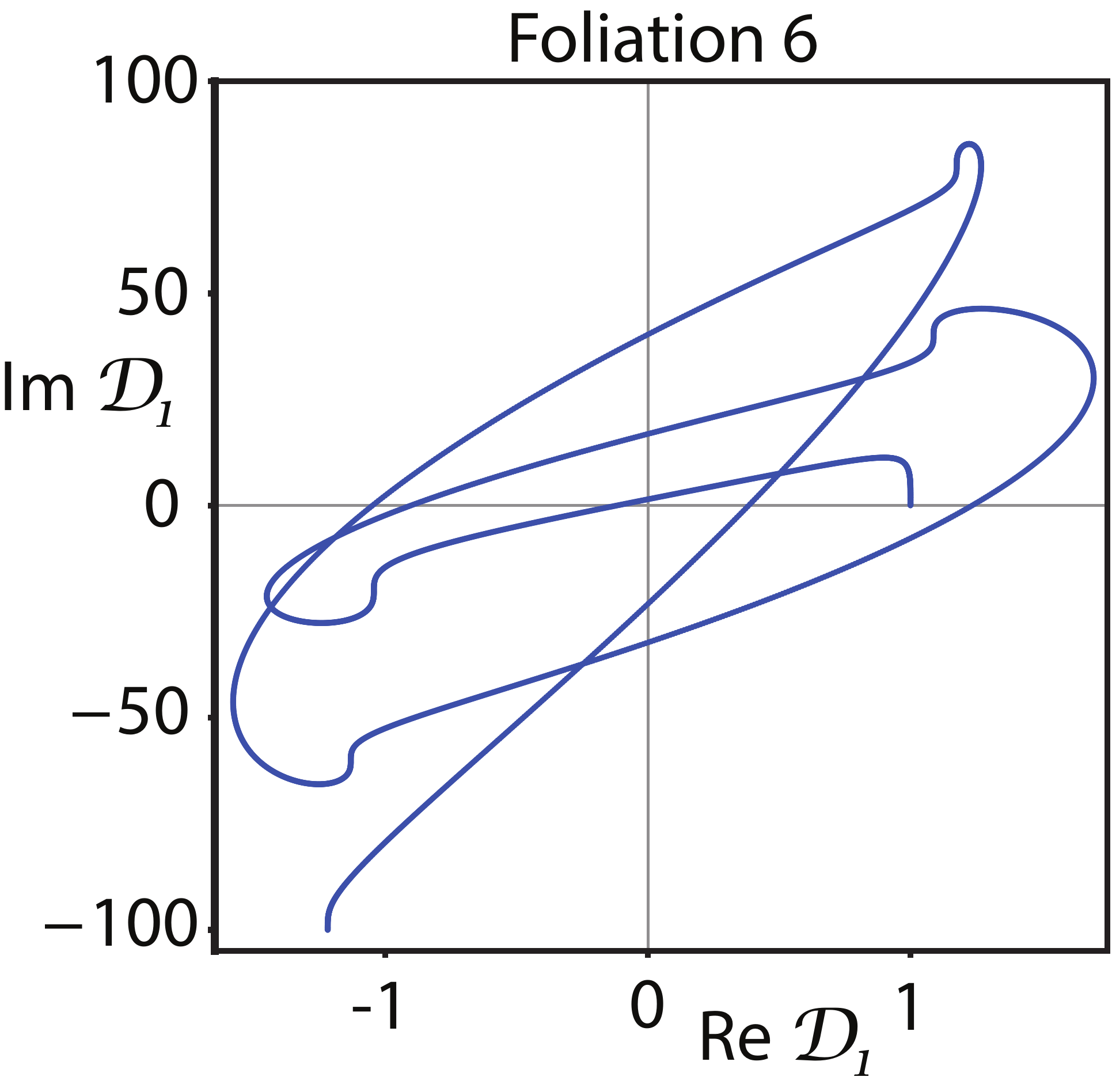}
\caption{Rotation sense of the total accumulated phase of the foliation \textcircled{\raisebox{-0.9pt}{6}} ($x=0$) determinant ${\cal D}_1$.  For roughly half a cycle of the motion it rotates counterclockwise, and then reverses itself to a clockwise rotation sense for the remainder of its propagation.  Nevertheless, its final point is very close to the final point of its associated real trajectory, which rotated exclusively counterclockwise.
 \label{fig8}} 
\end{figure}

As one would anticipate, it turns out that the stability matrices of the real trajectories and their associated saddles are very closely related.  In fact, their real parts are nearly identical.  See Fig.~\ref{fig6} for an illustration of the matrix elements contributing to ${\cal D}_1$.
Likewise, the imaginary parts of the saddle's stability matrix elements are very small at the final propagation time (to be compared to the real trajectory having no imaginary part), but at intermediate times can be a bit larger; see Fig.~\ref{fig7}.  In this case, the imaginary parts are approximately $2$ orders of magnitude smaller than the real 
parts at intermediate times, and even much smaller by the endpoint in time at $3\tau$.  Nevertheless, these small imaginary parts of $M_{22}$ and $M_{21}$ (it is multiplied by ${\bf b}_\alpha$) can be sufficient to reverse the rotation sense of ${\cal D}_1$ from counterclockwise (as for the real trajectory used to find the saddle) to clockwise.  Indeed this happens for foliation \textcircled{\raisebox{-0.9pt}{6}}.  Figure \ref{fig8} shows how it 
begins by rotating counterclockwise, but after half a cycle of motion reverses rotation sense for the remainder of the propagation.  It ends up very nearly at the same final point, as it must, but generates a Maslov index which differs from its related real trajectory by $8$.  Keeping in mind that the Maslov index modulo $4$ generates the same phase, the quite different indices in this case actually generate the same overall phase, but this can be considered an accident resulting from an even number of unwanted sign flips.  Foliation \textcircled{\raisebox{-0.9pt}{2}}'s real trajectory and complex trajectory generate Maslov indices $+1$ and $-1$, respectively, thus differ by $2$.  Foliation \textcircled{\raisebox{-0.9pt}{4}} generates $+3$ and $-3$, respectively, and thus differ by $6$. Hence, the signs given by the complex trajectories in these two foliations are incorrect.

An interesting feature for the quartic oscillator and the multi-site Bose-Hubbard model~\cite{Tomsovic18} is the fact that the problematic saddles always turned out to be related to energies below that of the Gaussian centroid's energy (calculating the energy using the real trajectories of the associated foliation).  This example was arranged so that the foliation \textcircled{\raisebox{-0.9pt}{7}} saddle at $x=0$ was real, not complex, and equivalent to the real trajectory used as a Newton-Raphson seed.  In Fig.~\ref{fig7}, it can be seen that the imaginary parts of the stability matrices start out in opposite directions for foliations \textcircled{\raisebox{-0.9pt}{6}} (goes negative first) and \textcircled{\raisebox{-0.9pt}{8}} (goes positive first).  Foliation \textcircled{\raisebox{-0.9pt}{8}} is associated with an energy greater than that of $(q_\alpha,p_\alpha)$ and all such foliations (at greater energies) have stability matrix elements that start in the same direction.  None of the determinants of these associated saddles ever reversed themselves and began winding clockwise.  On the other hand, all of the saddles associated with lower energies did and had Maslov indices that differed from their associated real trajectories.   This is consistent with the direction of motion of the zeros of ${\cal D}_1$ depicted in Fig.~\ref{fig4}.

In the quartic oscillator example, all the ${\cal D}_1$ determinant zeros lie above the $Im(t)=0$ axis for real trajectories.  Using the Maslov index derived from these trajectories automatically generates the same result as locating all zeros that have crossed the $Im(t)=0$ axis and deforming the time integration contour to pass beneath them.  This greatly simplifies the work by eliminating the necessity of locating the zeros of ${\cal D}_1$ (or ${\cal D}_2$) at complex times, and then creating deformed complex time integration contours.

\subsubsection{Hidden saddles}

Naturally, hidden saddles are those that cannot be located by using some real trajectory as a seed for a Newton-Raphson search.  Thus, the direct use of real trajectories does not apply as it does for the exposed saddles.  However, in~\cite{Wang21a} it was discussed that the best way to locate the physically relevant hidden saddles was to begin with exposed saddles (actually all of them for the quartic oscillator example), and use parameter variation to follow them continuously as they pass beyond caustic regions, the caustics being responsible to a large extent for the saddles being hidden to begin with.  Starting with an exposed saddle and a corrected Maslov index, one need only step finely through the parameter variation using a Newton-Raphson search to get from one saddle to the next until arriving at the desired hidden saddle.  By reversing any sign flips that may occur along the way, ensuring continuity, the Maslov indices for these hidden saddles end up with their corrected values.  Again, complex time is avoided and almost no extra calculations are needed beyond those required to locate the saddles, exposed or hidden.  The additional calculation necessary is just to monitor the determinants for sudden changes in total phase accumulation of $2\pi$, which effectively adds no extra computation time or complications beyond what already must exist.  

Finally, the remaining potentially, physically relevant saddles are those behind ``bald spots''~\cite{Petersen14}, which cannot be located with purely real time propagation.  Instead of the prefactor diverging because the determinant vanished, the prefactor itself vanishes for singular trajectories acquiring infinite momenta in finite times where the determinant diverges.  Again, that leads to an ambiguity in the phase choice even if one can locate the saddle.  Petersen and Kay showed that it was essential to deform the real time propagation path appropriately to complex time paths around the singular trajectories in order to arrive at these saddles~\cite{Petersen14} with the appropriate properties.  Nevertheless, just as discussed for the previous group of hidden saddles, it is not necessary to locate the migrating complex time zeros of the determinants for this class of hidden saddles either.  A combination of continuous parameter variation and complex time path deformation, which must be done in any case to locate them, suffices to fix any Maslov index ambiguities.

\section{Summary}
\label{sum}

The saddle point approximation applied to Glauber coherent states for bosonic many-body systems and wave packets in Schr\"odinger quantum mechanics is an important asymptotic analysis and approximation in a wide variety of physical contexts.  It always leads to the existence of a complex prefactor depending on the inverse square root of a determinant (labelled ${\cal D}_1$ and ${\cal D}_2$ here) for which the phase must be properly fixed.  The standard practice of: i) continuously following the total phase accumulation of the determinant's phase, ii) cutting it in half, and iii) inverting it, is insufficient from a theoretical perspective to fix the phase.  This prescription sometimes leads to sign errors.  The use of complex canonically conjugate variables (position, momentum) necessitates the complexification of time as well; see the work of~\cite{Petersen14} for the example of saddles behind bald spots and also note the use of complexification of time in solving tunneling problems~\cite{Creagh98}.  The determinants develop zeros in the complex time domain, which generate branch cuts in the prefactor.  The appropriate deformation of the real time integration path using a complex time path to circumvent the branch cuts resolves the issue and completes the theory.

It can be prohibitively difficult to locate complex time determinant zeros for systems with many degrees of freedom.  Instead, a practical way to evade their construction is given here based on a technique that borrows from the ideas behind the classification of saddles into exposed and hidden ones in the previous paper of this pair~\cite{Wang21a}.  Exposed saddles can be located by a Newton-Raphson technique with a real reference trajectory as a starting seed.  The real trajectory has a well defined Maslov index and there must be continuity between some particular saddle's contribution and Gaussian integration over its real seed trajectory.  The majority of the remaining hidden saddles can be found through parameter variation starting with exposed saddles, and ensuring continuity as the parameter changes is sufficient.  The calculations required for this practical approach entail almost no new calculations and are straightforward to implement.

The quartic oscillator used here to illustrate the sign problem and its resolution is homogeneous.  Its dynamics are especially straightforward.  It would be good to examine the practical solution in more complicated dynamical situations, including systems with many degrees of freedom.  For example, the existence of saddles behind bald spots~\cite{Petersen14} does not exist for the quartic oscillator, and these saddles require both parameter variation and complex time paths to discover and treat.  Thus, some saddles require the use of complex time for generating trajectories on the correct sheets. The practical resolution of the phase index problem outlined in this work (continuous parameter variation along with requiring continuity of the phase) continues to hold nevertheless even for these more difficult-to-locate saddles.

\section{Acknowledgments}
We would like to thank Dr. Stephen Creagh from the University of Nottingham for insightful discussions. He suggested that complex time may play a fundamental role.

\bibliography{quantumchaos,manybody,general_ref,molecular,rmtmodify}

\begin{thebibliography}{37}%
\makeatletter
\providecommand \@ifxundefined [1]{%
 \@ifx{#1\undefined}
}%
\providecommand \@ifnum [1]{%
 \ifnum #1\expandafter \@firstoftwo
 \else \expandafter \@secondoftwo
 \fi
}%
\providecommand \@ifx [1]{%
 \ifx #1\expandafter \@firstoftwo
 \else \expandafter \@secondoftwo
 \fi
}%
\providecommand \natexlab [1]{#1}%
\providecommand \enquote  [1]{``#1''}%
\providecommand \bibnamefont  [1]{#1}%
\providecommand \bibfnamefont [1]{#1}%
\providecommand \citenamefont [1]{#1}%
\providecommand \href@noop [0]{\@secondoftwo}%
\providecommand \href [0]{\begingroup \@sanitize@url \@href}%
\providecommand \@href[1]{\@@startlink{#1}\@@href}%
\providecommand \@@href[1]{\endgroup#1\@@endlink}%
\providecommand \@sanitize@url [0]{\catcode `\\12\catcode `\$12\catcode
  `\&12\catcode `\#12\catcode `\^12\catcode `\_12\catcode `\%12\relax}%
\providecommand \@@startlink[1]{}%
\providecommand \@@endlink[0]{}%
\providecommand \url  [0]{\begingroup\@sanitize@url \@url }%
\providecommand \@url [1]{\endgroup\@href {#1}{\urlprefix }}%
\providecommand \urlprefix  [0]{URL }%
\providecommand \Eprint [0]{\href }%
\providecommand \doibase [0]{https://doi.org/}%
\providecommand \selectlanguage [0]{\@gobble}%
\providecommand \bibinfo  [0]{\@secondoftwo}%
\providecommand \bibfield  [0]{\@secondoftwo}%
\providecommand \translation [1]{[#1]}%
\providecommand \BibitemOpen [0]{}%
\providecommand \bibitemStop [0]{}%
\providecommand \bibitemNoStop [0]{.\EOS\space}%
\providecommand \EOS [0]{\spacefactor3000\relax}%
\providecommand \BibitemShut  [1]{\csname bibitem#1\endcsname}%
\let\auto@bib@innerbib\@empty
%</preamble>
\bibitem [{\citenamefont {Einstein}(1917)}]{Einstein17}%
  \BibitemOpen
  \bibfield  {author} {\bibinfo {author} {\bibfnamefont {A.}~\bibnamefont
  {Einstein}},\ }\bibfield  {title} {\bibinfo {title} {Zum quantensatz von
  sommerfeld und epstein},\ }\href@noop {} {\bibfield  {journal} {\bibinfo
  {journal} {Verh.~Deutsch.~Phys.~Ges.~Berlin}\ }\textbf {\bibinfo {volume}
  {19}},\ \bibinfo {pages} {82} (\bibinfo {year} {1917})},\ \bibinfo {note}
  {english translation by C.~Jaffe, JILA report no.~116, 1980}\BibitemShut
  {NoStop}%
\bibitem [{\citenamefont {Jeffreys}(1924)}]{Jeffreys24}%
  \BibitemOpen
  \bibfield  {author} {\bibinfo {author} {\bibfnamefont {H.}~\bibnamefont
  {Jeffreys}},\ }\bibfield  {title} {\bibinfo {title} {On certain approximate
  solutions of linear differential equations of the second order},\ }\href@noop
  {} {\bibfield  {journal} {\bibinfo  {journal} {Proc.~of the
  Lond.~Math.~Soc.}\ }\textbf {\bibinfo {volume} {23}},\ \bibinfo {pages} {428}
  (\bibinfo {year} {1924})}\BibitemShut {NoStop}%
\bibitem [{\citenamefont {Wentzel}(1926)}]{Wentzel26}%
  \BibitemOpen
  \bibfield  {author} {\bibinfo {author} {\bibfnamefont {G.}~\bibnamefont
  {Wentzel}},\ }\bibfield  {title} {\bibinfo {title} {Eine verallgemeinerung
  der quantenbedingungen f\"uŸr die zwecke der wellenmechanik},\ }\href@noop {}
  {\bibfield  {journal} {\bibinfo  {journal} {Zeit.~der Phys.}\ }\textbf
  {\bibinfo {volume} {38}},\ \bibinfo {pages} {518} (\bibinfo {year}
  {1926})}\BibitemShut {NoStop}%
\bibitem [{\citenamefont {Kramers}(1926)}]{Kramers26}%
  \BibitemOpen
  \bibfield  {author} {\bibinfo {author} {\bibfnamefont {H.~A.}\ \bibnamefont
  {Kramers}},\ }\bibfield  {title} {\bibinfo {title} {Wellenmechanik und
  halbzŠhlige quantisierung},\ }\href@noop {} {\bibfield  {journal} {\bibinfo
  {journal} {Zeit.~der Phys.}\ }\textbf {\bibinfo {volume} {39}},\ \bibinfo
  {pages} {828} (\bibinfo {year} {1926})}\BibitemShut {NoStop}%
\bibitem [{\citenamefont {Brillouin}(1926)}]{Brillouin26}%
  \BibitemOpen
  \bibfield  {author} {\bibinfo {author} {\bibfnamefont {L.}~\bibnamefont
  {Brillouin}},\ }\bibfield  {title} {\bibinfo {title} {La m\'ecanique
  ondulatoire de schr\"odinger: une mŽthode g\'en\'erale de resolution par
  approximations successives},\ }\href@noop {} {\bibfield  {journal} {\bibinfo
  {journal} {Comptes Rendus de l'Acad.~des Sci.}\ }\textbf {\bibinfo {volume}
  {183}},\ \bibinfo {pages} {24} (\bibinfo {year} {1926})}\BibitemShut
  {NoStop}%
\bibitem [{\citenamefont {Keller}(1958)}]{Keller58}%
  \BibitemOpen
  \bibfield  {author} {\bibinfo {author} {\bibfnamefont {J.~B.}\ \bibnamefont
  {Keller}},\ }\bibfield  {title} {\bibinfo {title} {Corrected bohr-sommerfeld
  quantum conditions for nonseparable systems},\ }\href@noop {} {\bibfield
  {journal} {\bibinfo  {journal} {Ann.~Phys. (N.Y.)}\ }\textbf {\bibinfo
  {volume} {4}},\ \bibinfo {pages} {180} (\bibinfo {year} {1958})}\BibitemShut
  {NoStop}%
\bibitem [{\citenamefont {Maslov}\ and\ \citenamefont
  {Fedoriuk}(1981)}]{Maslov81}%
  \BibitemOpen
  \bibfield  {author} {\bibinfo {author} {\bibfnamefont {V.~P.}\ \bibnamefont
  {Maslov}}\ and\ \bibinfo {author} {\bibfnamefont {M.~V.}\ \bibnamefont
  {Fedoriuk}},\ }\href@noop {} {\emph {\bibinfo {title} {Semiclassical
  approximation in quantum mechanics}}}\ (\bibinfo  {publisher} {Reidel
  Publishing Company},\ \bibinfo {address} {Dordrecht},\ \bibinfo {year}
  {1981})\BibitemShut {NoStop}%
\bibitem [{\citenamefont {Klauder}\ and\ \citenamefont
  {Skagerstam}(1985)}]{Klauder85}%
  \BibitemOpen
  \bibfield  {author} {\bibinfo {author} {\bibfnamefont {J.~R.}\ \bibnamefont
  {Klauder}}\ and\ \bibinfo {author} {\bibfnamefont {B.-S.}\ \bibnamefont
  {Skagerstam}},\ }\href@noop {} {\emph {\bibinfo {title} {Coherent States:
  Applications in Physics and Mathematical Physics}}}\ (\bibinfo  {publisher}
  {World Scientific},\ \bibinfo {address} {Singapore},\ \bibinfo {year}
  {1985})\BibitemShut {NoStop}%
\bibitem [{\citenamefont {Baranger}\ \emph {et~al.}(2001)\citenamefont
  {Baranger}, \citenamefont {de~Aguiar}, \citenamefont {Keck}, \citenamefont
  {Korsch},\ and\ \citenamefont {Schellhaass}}]{Baranger01}%
  \BibitemOpen
  \bibfield  {author} {\bibinfo {author} {\bibfnamefont {M.}~\bibnamefont
  {Baranger}}, \bibinfo {author} {\bibfnamefont {M.~A.~M.}\ \bibnamefont
  {de~Aguiar}}, \bibinfo {author} {\bibfnamefont {F.}~\bibnamefont {Keck}},
  \bibinfo {author} {\bibfnamefont {H.~J.}\ \bibnamefont {Korsch}},\ and\
  \bibinfo {author} {\bibfnamefont {B.}~\bibnamefont {Schellhaass}},\
  }\bibfield  {title} {\bibinfo {title} {Semiclassical approximations in phase
  space with coherent states},\ }\href@noop {} {\bibfield  {journal} {\bibinfo
  {journal} {J.~Phys.~A:~Math.~Gen.}\ }\textbf {\bibinfo {volume} {34}},\
  \bibinfo {pages} {7227} (\bibinfo {year} {2001})}\BibitemShut {NoStop}%
\bibitem [{\citenamefont {Engl}\ \emph {et~al.}(2015)\citenamefont {Engl},
  \citenamefont {Urbina},\ and\ \citenamefont {Richter}}]{Engl15}%
  \BibitemOpen
  \bibfield  {author} {\bibinfo {author} {\bibfnamefont {T.}~\bibnamefont
  {Engl}}, \bibinfo {author} {\bibfnamefont {J.~D.}\ \bibnamefont {Urbina}},\
  and\ \bibinfo {author} {\bibfnamefont {K.}~\bibnamefont {Richter}},\
  }\bibfield  {title} {\bibinfo {title} {Periodic mean-field solutions and the
  spectra of discrete bosonic fields: Trace formula for bose-hubbard models},\
  }\href@noop {} {\bibfield  {journal} {\bibinfo  {journal} {Phys.~Rev.~E}\
  }\textbf {\bibinfo {volume} {92}},\ \bibinfo {pages} {062907} (\bibinfo
  {year} {2015})}\BibitemShut {NoStop}%
\bibitem [{\citenamefont {Gutzwiller}(1971)}]{Gutzwiller71}%
  \BibitemOpen
  \bibfield  {author} {\bibinfo {author} {\bibfnamefont {M.~C.}\ \bibnamefont
  {Gutzwiller}},\ }\bibfield  {title} {\bibinfo {title} {Periodic orbits and
  classical quantization conditions},\ }\href@noop {} {\bibfield  {journal}
  {\bibinfo  {journal} {J.~Math.~Phys.}\ }\textbf {\bibinfo {volume} {12}},\
  \bibinfo {pages} {343} (\bibinfo {year} {1971})},\ \bibinfo {note} {and
  references therein}\BibitemShut {NoStop}%
\bibitem [{\citenamefont {Creagh}\ \emph {et~al.}(1990)\citenamefont {Creagh},
  \citenamefont {Robbins},\ and\ \citenamefont {Littlejohn}}]{Creagh90}%
  \BibitemOpen
  \bibfield  {author} {\bibinfo {author} {\bibfnamefont {S.~C.}\ \bibnamefont
  {Creagh}}, \bibinfo {author} {\bibfnamefont {J.~M.}\ \bibnamefont
  {Robbins}},\ and\ \bibinfo {author} {\bibfnamefont {R.~G.}\ \bibnamefont
  {Littlejohn}},\ }\bibfield  {title} {\bibinfo {title} {Geometrical properties
  of maslov indices in the semiclassical trace formula for the density of
  states},\ }\href@noop {} {\bibfield  {journal} {\bibinfo  {journal}
  {Phys.~Rev.~A}\ }\textbf {\bibinfo {volume} {42}},\ \bibinfo {pages} {1907}
  (\bibinfo {year} {1990})}\BibitemShut {NoStop}%
\bibitem [{\citenamefont {Vleck}(1928)}]{Vanvleck28}%
  \BibitemOpen
  \bibfield  {author} {\bibinfo {author} {\bibfnamefont {J.~H.~V.}\
  \bibnamefont {Vleck}},\ }\bibfield  {title} {\bibinfo {title} {The
  correspondence principle in the statistical interpretation of quantum
  mechanics},\ }\href@noop {} {\bibfield  {journal} {\bibinfo  {journal}
  {Proc.~Natl.~Acad.~Sci.~U.S.A.}\ }\textbf {\bibinfo {volume} {14}},\ \bibinfo
  {pages} {178} (\bibinfo {year} {1928})}\BibitemShut {NoStop}%
\bibitem [{\citenamefont {Schulman}(1981)}]{Schulman81}%
  \BibitemOpen
  \bibfield  {author} {\bibinfo {author} {\bibfnamefont {L.~S.}\ \bibnamefont
  {Schulman}},\ }\href@noop {} {\emph {\bibinfo {title} {Techniques and
  Applications of Path Integration}}}\ (\bibinfo  {publisher} {John Wiley and
  Sons, Inc.},\ \bibinfo {address} {New York},\ \bibinfo {year}
  {1981})\BibitemShut {NoStop}%
\bibitem [{\citenamefont {Littlejohn}(1986)}]{Littlejohn86}%
  \BibitemOpen
  \bibfield  {author} {\bibinfo {author} {\bibfnamefont {R.~G.}\ \bibnamefont
  {Littlejohn}},\ }\bibfield  {title} {\bibinfo {title} {The semiclassical
  evolution of wave packets},\ }\href@noop {} {\bibfield  {journal} {\bibinfo
  {journal} {Phys.~Rep.}\ }\textbf {\bibinfo {volume} {138}},\ \bibinfo {pages}
  {193} (\bibinfo {year} {1986})}\BibitemShut {NoStop}%
\bibitem [{\citenamefont {Glauber}(1963)}]{Glauber63}%
  \BibitemOpen
  \bibfield  {author} {\bibinfo {author} {\bibfnamefont {R.~J.}\ \bibnamefont
  {Glauber}},\ }\bibfield  {title} {\bibinfo {title} {Coherent and incoherent
  states of the radiation field},\ }\href@noop {} {\bibfield  {journal}
  {\bibinfo  {journal} {Phys.~Rev.}\ }\textbf {\bibinfo {volume} {131}},\
  \bibinfo {pages} {2766} (\bibinfo {year} {1963})}\BibitemShut {NoStop}%
\bibitem [{\citenamefont {Huber}\ \emph {et~al.}(1988)\citenamefont {Huber},
  \citenamefont {Heller},\ and\ \citenamefont {Littlejohn}}]{Huber88}%
  \BibitemOpen
  \bibfield  {author} {\bibinfo {author} {\bibfnamefont {D.}~\bibnamefont
  {Huber}}, \bibinfo {author} {\bibfnamefont {E.~J.}\ \bibnamefont {Heller}},\
  and\ \bibinfo {author} {\bibfnamefont {R.~G.}\ \bibnamefont {Littlejohn}},\
  }\bibfield  {title} {\bibinfo {title} {Generalized gaussian wave packet
  dynamics, schr\"odinger equation, and stationary phase approximation},\
  }\href@noop {} {\bibfield  {journal} {\bibinfo  {journal} {J.~Chem.~Phys.}\
  }\textbf {\bibinfo {volume} {89}},\ \bibinfo {pages} {2003} (\bibinfo {year}
  {1988})}\BibitemShut {NoStop}%
\bibitem [{\citenamefont {Scully}\ and\ \citenamefont
  {Zubairy}(1997)}]{Scully97}%
  \BibitemOpen
  \bibfield  {author} {\bibinfo {author} {\bibfnamefont {M.~O.}\ \bibnamefont
  {Scully}}\ and\ \bibinfo {author} {\bibfnamefont {M.~S.}\ \bibnamefont
  {Zubairy}},\ }\href@noop {} {\emph {\bibinfo {title} {Quantum Optics}}}\
  (\bibinfo  {publisher} {Cambridge University Press},\ \bibinfo {address}
  {Cambridge, UK},\ \bibinfo {year} {1997})\BibitemShut {NoStop}%
\bibitem [{\citenamefont {Greiner}\ \emph {et~al.}(2002)\citenamefont
  {Greiner}, \citenamefont {Mandel}, \citenamefont {H\"ansch},\ and\
  \citenamefont {Bloch}}]{Greiner02b}%
  \BibitemOpen
  \bibfield  {author} {\bibinfo {author} {\bibfnamefont {M.}~\bibnamefont
  {Greiner}}, \bibinfo {author} {\bibfnamefont {O.}~\bibnamefont {Mandel}},
  \bibinfo {author} {\bibfnamefont {T.~W.}\ \bibnamefont {H\"ansch}},\ and\
  \bibinfo {author} {\bibfnamefont {I.}~\bibnamefont {Bloch}},\ }\bibfield
  {title} {\bibinfo {title} {Collapse and revival of the matter wave field of a
  bose-einstein condensate},\ }\href@noop {} {\bibfield  {journal} {\bibinfo
  {journal} {Nature}\ }\textbf {\bibinfo {volume} {419}},\ \bibinfo {pages}
  {51} (\bibinfo {year} {2002})}\BibitemShut {NoStop}%
\bibitem [{\citenamefont {Polkovnikov}\ \emph {et~al.}(2011)\citenamefont
  {Polkovnikov}, \citenamefont {Sengupta}, \citenamefont {Silva},\ and\
  \citenamefont {Vengalattore}}]{Polkovnikov11}%
  \BibitemOpen
  \bibfield  {author} {\bibinfo {author} {\bibfnamefont {A.}~\bibnamefont
  {Polkovnikov}}, \bibinfo {author} {\bibfnamefont {K.}~\bibnamefont
  {Sengupta}}, \bibinfo {author} {\bibfnamefont {A.}~\bibnamefont {Silva}},\
  and\ \bibinfo {author} {\bibfnamefont {M.}~\bibnamefont {Vengalattore}},\
  }\bibfield  {title} {\bibinfo {title} {Colloquium: Nonequilibrium dynamics of
  closed interacting quantum systems},\ }\href@noop {} {\bibfield  {journal}
  {\bibinfo  {journal} {Rev.~Mod.~Phys.}\ }\textbf {\bibinfo {volume} {83}},\
  \bibinfo {pages} {863} (\bibinfo {year} {2011})}\BibitemShut {NoStop}%
\bibitem [{\citenamefont {Heller}(1981)}]{Heller81b}%
  \BibitemOpen
  \bibfield  {author} {\bibinfo {author} {\bibfnamefont {E.~J.}\ \bibnamefont
  {Heller}},\ }\bibfield  {title} {\bibinfo {title} {The semiclassical way to
  molecular spectroscopy},\ }\href@noop {} {\bibfield  {journal} {\bibinfo
  {journal} {Acc.~Chem.~Res.}\ }\textbf {\bibinfo {volume} {14}},\ \bibinfo
  {pages} {368} (\bibinfo {year} {1981})}\BibitemShut {NoStop}%
\bibitem [{\citenamefont {Gruebele}\ and\ \citenamefont
  {Zewail}(1992)}]{Gruebele92}%
  \BibitemOpen
  \bibfield  {author} {\bibinfo {author} {\bibfnamefont {M.}~\bibnamefont
  {Gruebele}}\ and\ \bibinfo {author} {\bibfnamefont {A.~H.}\ \bibnamefont
  {Zewail}},\ }\bibfield  {title} {\bibinfo {title} {Femtosecond wave packet
  spectroscopy: Coherences, the potential, and structural determination},\
  }\href@noop {} {\bibfield  {journal} {\bibinfo  {journal} {J.~Chem.~Phys.}\
  }\textbf {\bibinfo {volume} {98}},\ \bibinfo {pages} {883} (\bibinfo {year}
  {1992})}\BibitemShut {NoStop}%
\bibitem [{\citenamefont {Zewail}(2000)}]{Zewail00}%
  \BibitemOpen
  \bibfield  {author} {\bibinfo {author} {\bibfnamefont {A.~H.}\ \bibnamefont
  {Zewail}},\ }\bibfield  {title} {\bibinfo {title} {Femtochemistry:
  Atomic-scale dynamics of the chemical bond},\ }\href@noop {} {\bibfield
  {journal} {\bibinfo  {journal} {J.~Phys.~Chem.~A}\ }\textbf {\bibinfo
  {volume} {104}},\ \bibinfo {pages} {5660} (\bibinfo {year}
  {2000})}\BibitemShut {NoStop}%
\bibitem [{\citenamefont {Agostini}\ and\ \citenamefont
  {DiMauro}(2004)}]{Agostini04}%
  \BibitemOpen
  \bibfield  {author} {\bibinfo {author} {\bibfnamefont {P.}~\bibnamefont
  {Agostini}}\ and\ \bibinfo {author} {\bibfnamefont {L.~F.}\ \bibnamefont
  {DiMauro}},\ }\bibfield  {title} {\bibinfo {title} {The physics of attosecond
  light pulses},\ }\href@noop {} {\bibfield  {journal} {\bibinfo  {journal}
  {Rep.~Prog.~Phys.}\ }\textbf {\bibinfo {volume} {67}},\ \bibinfo {pages}
  {813} (\bibinfo {year} {2004})}\BibitemShut {NoStop}%
\bibitem [{\citenamefont {Tomsovic}\ \emph {et~al.}(2018)\citenamefont
  {Tomsovic}, \citenamefont {Schlagheck}, \citenamefont {Ullmo}, \citenamefont
  {Urbina},\ and\ \citenamefont {Richter}}]{Tomsovic18}%
  \BibitemOpen
  \bibfield  {author} {\bibinfo {author} {\bibfnamefont {S.}~\bibnamefont
  {Tomsovic}}, \bibinfo {author} {\bibfnamefont {P.}~\bibnamefont
  {Schlagheck}}, \bibinfo {author} {\bibfnamefont {D.}~\bibnamefont {Ullmo}},
  \bibinfo {author} {\bibfnamefont {J.-D.}\ \bibnamefont {Urbina}},\ and\
  \bibinfo {author} {\bibfnamefont {K.}~\bibnamefont {Richter}},\ }\bibfield
  {title} {\bibinfo {title} {Post-ehrenfest many-body quantum interferences in
  ultracold atoms far-out-of-equilibrium},\ }\href@noop {} {\bibfield
  {journal} {\bibinfo  {journal} {Phys.~Rev.~A}\ }\textbf {\bibinfo {volume}
  {97}},\ \bibinfo {pages} {061606(R)} (\bibinfo {year} {2018})},\ \bibinfo
  {note} {arXiv:1711.04693v2 [quant-ph]}\BibitemShut {NoStop}%
\bibitem [{\citenamefont {Wang}\ and\ \citenamefont
  {Tomsovic}(2021)}]{Wang21a}%
  \BibitemOpen
  \bibfield  {author} {\bibinfo {author} {\bibfnamefont {H.}~\bibnamefont
  {Wang}}\ and\ \bibinfo {author} {\bibfnamefont {S.}~\bibnamefont
  {Tomsovic}},\ }\bibfield  {title} {\bibinfo {title} {Semiclassical
  propagation of coherent states and wave packets: hidden saddles},\
  }\href@noop {} {\bibfield  {journal} {\bibinfo  {journal} {arXiv:2107.08799
  [quant-ph]}\ } (\bibinfo {year} {2021})}\BibitemShut {NoStop}%
\bibitem [{\citenamefont {Tomsovic}(2018)}]{Tomsovic18b}%
  \BibitemOpen
  \bibfield  {author} {\bibinfo {author} {\bibfnamefont {S.}~\bibnamefont
  {Tomsovic}},\ }\bibfield  {title} {\bibinfo {title} {Complex saddle
  trajectories for multidimensional quantum wave packet/coherent state
  propagation: application to a many-body system},\ }\href@noop {} {\bibfield
  {journal} {\bibinfo  {journal} {Phys.~Rev.~E}\ }\textbf {\bibinfo {volume}
  {98}},\ \bibinfo {pages} {023301} (\bibinfo {year} {2018})},\ \bibinfo {note}
  {arXiv:1804.10511 [cond-mat.stat-mech]}\BibitemShut {NoStop}%
\bibitem [{\citenamefont {Bohigas}\ \emph {et~al.}(1993)\citenamefont
  {Bohigas}, \citenamefont {Tomsovic},\ and\ \citenamefont
  {Ullmo}}]{Bohigas93}%
  \BibitemOpen
  \bibfield  {author} {\bibinfo {author} {\bibfnamefont {O.}~\bibnamefont
  {Bohigas}}, \bibinfo {author} {\bibfnamefont {S.}~\bibnamefont {Tomsovic}},\
  and\ \bibinfo {author} {\bibfnamefont {D.}~\bibnamefont {Ullmo}},\ }\bibfield
   {title} {\bibinfo {title} {Manifestations of classical phase space
  structures in quantum mechanics},\ }\href@noop {} {\bibfield  {journal}
  {\bibinfo  {journal} {Phys.~Rep.}\ }\textbf {\bibinfo {volume} {223}},\
  \bibinfo {pages} {43} (\bibinfo {year} {1993})}\BibitemShut {NoStop}%
\bibitem [{\citenamefont {Van~Voorhis}\ and\ \citenamefont
  {Heller}(2002)}]{Vanvoorhis02}%
  \BibitemOpen
  \bibfield  {author} {\bibinfo {author} {\bibfnamefont {T.}~\bibnamefont
  {Van~Voorhis}}\ and\ \bibinfo {author} {\bibfnamefont {E.~J.}\ \bibnamefont
  {Heller}},\ }\bibfield  {title} {\bibinfo {title} {Nearly real trajectories
  in complex semiclassical dynamics},\ }\href@noop {} {\bibfield  {journal}
  {\bibinfo  {journal} {Phys.~Rev.~A}\ }\textbf {\bibinfo {volume} {66}},\
  \bibinfo {pages} {050501(R)} (\bibinfo {year} {2002})}\BibitemShut {NoStop}%
\bibitem [{\citenamefont {Van~Voorhis}\ and\ \citenamefont
  {Heller}(2003)}]{Vanvoorhis03}%
  \BibitemOpen
  \bibfield  {author} {\bibinfo {author} {\bibfnamefont {T.}~\bibnamefont
  {Van~Voorhis}}\ and\ \bibinfo {author} {\bibfnamefont {E.~J.}\ \bibnamefont
  {Heller}},\ }\bibfield  {title} {\bibinfo {title} {Similarity transformed
  semiclassical dynamics},\ }\href@noop {} {\bibfield  {journal} {\bibinfo
  {journal} {J.~Chem.~Phys.}\ }\textbf {\bibinfo {volume} {119}},\ \bibinfo
  {pages} {12153} (\bibinfo {year} {2003})}\BibitemShut {NoStop}%
\bibitem [{\citenamefont {Pal}\ \emph {et~al.}(2016)\citenamefont {Pal},
  \citenamefont {Vyas},\ and\ \citenamefont {Tomsovic}}]{Pal16}%
  \BibitemOpen
  \bibfield  {author} {\bibinfo {author} {\bibfnamefont {H.}~\bibnamefont
  {Pal}}, \bibinfo {author} {\bibfnamefont {M.}~\bibnamefont {Vyas}},\ and\
  \bibinfo {author} {\bibfnamefont {S.}~\bibnamefont {Tomsovic}},\ }\bibfield
  {title} {\bibinfo {title} {Generalized gaussian wave packet dynamics:
  Integrable and chaotic systems},\ }\href@noop {} {\bibfield  {journal}
  {\bibinfo  {journal} {Phys.~Rev.~E}\ }\textbf {\bibinfo {volume} {93}},\
  \bibinfo {pages} {012213} (\bibinfo {year} {2016})},\ \bibinfo {note}
  {arXiv:1510.08051 [quant-ph]}\BibitemShut {NoStop}%
\bibitem [{\citenamefont {O'Connor}\ \emph {et~al.}(1992)\citenamefont
  {O'Connor}, \citenamefont {Tomsovic},\ and\ \citenamefont
  {Heller}}]{Oconnor92}%
  \BibitemOpen
  \bibfield  {author} {\bibinfo {author} {\bibfnamefont {P.~W.}\ \bibnamefont
  {O'Connor}}, \bibinfo {author} {\bibfnamefont {S.}~\bibnamefont {Tomsovic}},\
  and\ \bibinfo {author} {\bibfnamefont {E.~J.}\ \bibnamefont {Heller}},\
  }\bibfield  {title} {\bibinfo {title} {Semiclassical dynamics in the strongly
  chaotic regime - breaking the log-time barrier},\ }\href@noop {} {\bibfield
  {journal} {\bibinfo  {journal} {Physica D}\ }\textbf {\bibinfo {volume}
  {55}},\ \bibinfo {pages} {340} (\bibinfo {year} {1992})}\BibitemShut
  {NoStop}%
\bibitem [{\citenamefont {Tomsovic}\ and\ \citenamefont
  {Heller}(1993{\natexlab{a}})}]{Tomsovic93}%
  \BibitemOpen
  \bibfield  {author} {\bibinfo {author} {\bibfnamefont {S.}~\bibnamefont
  {Tomsovic}}\ and\ \bibinfo {author} {\bibfnamefont {E.~J.}\ \bibnamefont
  {Heller}},\ }\bibfield  {title} {\bibinfo {title} {The long-time
  semiclassical dynamics of chaos: the stadium billiard},\ }\href@noop {}
  {\bibfield  {journal} {\bibinfo  {journal} {Phys.~Rev.~E}\ }\textbf {\bibinfo
  {volume} {47}},\ \bibinfo {pages} {282} (\bibinfo {year}
  {1993}{\natexlab{a}})}\BibitemShut {NoStop}%
\bibitem [{\citenamefont {Tomsovic}\ and\ \citenamefont
  {Heller}(1993{\natexlab{b}})}]{Tomsovic93b}%
  \BibitemOpen
  \bibfield  {author} {\bibinfo {author} {\bibfnamefont {S.}~\bibnamefont
  {Tomsovic}}\ and\ \bibinfo {author} {\bibfnamefont {E.~J.}\ \bibnamefont
  {Heller}},\ }\bibfield  {title} {\bibinfo {title} {The semiclassical
  construction of chaotic eigenstates},\ }\href@noop {} {\bibfield  {journal}
  {\bibinfo  {journal} {Phys.~Rev.~Lett.}\ }\textbf {\bibinfo {volume} {70}},\
  \bibinfo {pages} {1405} (\bibinfo {year} {1993}{\natexlab{b}})}\BibitemShut
  {NoStop}%
\bibitem [{\citenamefont {Barnes}\ \emph {et~al.}(1994)\citenamefont {Barnes},
  \citenamefont {Nauenberg}, \citenamefont {Nockleby},\ and\ \citenamefont
  {Tomsovic}}]{Barnes94}%
  \BibitemOpen
  \bibfield  {author} {\bibinfo {author} {\bibfnamefont {I.~M.~S.}\
  \bibnamefont {Barnes}}, \bibinfo {author} {\bibfnamefont {M.}~\bibnamefont
  {Nauenberg}}, \bibinfo {author} {\bibfnamefont {M.}~\bibnamefont
  {Nockleby}},\ and\ \bibinfo {author} {\bibfnamefont {S.}~\bibnamefont
  {Tomsovic}},\ }\bibfield  {title} {\bibinfo {title} {Classical orbits and
  semiclassical wave packet propagation in the coulomb potential},\ }\href@noop
  {} {\bibfield  {journal} {\bibinfo  {journal} {J.~Phys.~A: Math.~Gen.}\
  }\textbf {\bibinfo {volume} {27}},\ \bibinfo {pages} {3299} (\bibinfo {year}
  {1994})}\BibitemShut {NoStop}%
\bibitem [{\citenamefont {Petersen}\ and\ \citenamefont
  {Kay}(2014)}]{Petersen14}%
  \BibitemOpen
  \bibfield  {author} {\bibinfo {author} {\bibfnamefont {J.}~\bibnamefont
  {Petersen}}\ and\ \bibinfo {author} {\bibfnamefont {K.~G.}\ \bibnamefont
  {Kay}},\ }\bibfield  {title} {\bibinfo {title} {Complex time paths for
  semiclassical wave packet propagation with complex trajectories},\
  }\href@noop {} {\bibfield  {journal} {\bibinfo  {journal} {J.~Chem.~Phys.}\
  }\textbf {\bibinfo {volume} {141}},\ \bibinfo {pages} {054114} (\bibinfo
  {year} {2014})}\BibitemShut {NoStop}%
\bibitem [{\citenamefont {Creagh}(1998)}]{Creagh98}%
  \BibitemOpen
  \bibfield  {author} {\bibinfo {author} {\bibfnamefont {S.~C.}\ \bibnamefont
  {Creagh}},\ }\bibfield  {title} {\bibinfo {title} {Tunneling in two
  dimensions},\ }in\ \href@noop {} {\emph {\bibinfo {booktitle} {Tunneling in
  complex systems, Proceedings from the Institute for Nuclear Theory: Volume
  5}}},\ \bibinfo {editor} {edited by\ \bibinfo {editor} {\bibfnamefont
  {S.}~\bibnamefont {Tomsovic}}}\ (\bibinfo  {publisher} {World Scientific},\
  \bibinfo {address} {Singapore},\ \bibinfo {year} {1998})\ pp.\ \bibinfo
  {pages} {35--100}\BibitemShut {NoStop}%
\end{thebibliography}%

\end{document}